\documentclass[conference]{IEEEtran}
\IEEEoverridecommandlockouts
\usepackage{cite}
\usepackage{amsmath,amssymb,amsfonts}
\usepackage{algorithmic}
\usepackage{graphicx}
\usepackage{epstopdf}
\usepackage{textcomp}
\usepackage{xcolor}
\usepackage{comment}
\usepackage[caption=false]{subfig}
\def\BibTeX{{\rm B\kern-.05em{\sc i\kern-.025em b}\kern-.08em
    T\kern-.1667em\lower.7ex\hbox{E}\kern-.125emX}}

\begin{document}

\title{Optimization of Operation Strategy for \\Primary Torque based hydrostatic Drivetrain \\ using  Artificial Intelligence\\
}
 
\author{\IEEEauthorblockN{1\textsuperscript{st} Yusheng Xiang}  
\IEEEauthorblockA{\textit{Institute of  Vehicle System Technology} \\
\textit{Karlsruhe Institute of Technology}\\
Karlsruhe, Germany \\
E-mail adress: yusheng.xiang@partner.kit.edu}
\and
\IEEEauthorblockN{2\textsuperscript{nd} Marcus Geimer}
\IEEEauthorblockA{\textit{Institute of  Vehicle System Technology} \\
\textit{Karlsruhe Institute of Technology}\\
Karlsruhe, Germany \\
E-mail adress: marcus.geimer@kit.edu}

}

\maketitle

\begin{abstract}
A new primary torque control concept for hydrostatics mobile machines was introduced in 2018 \cite{rMutschler.2018}. The mentioned concept controls the pressure in a closed circuit by changing the angle of the hydraulic pump to achieve the desired pressure based on a feedback system. Thanks to this concept, a series of advantages are expected \cite{Xiang.2020}. However, while working in a Y cycle, the primary torque controlled wheel loader has worse performance in efficiency compared to secondary controlled earthmover due to lack of recuperation ability.  Alternatively, we use deep learning algorithms to improve machines’ regeneration performance. In this paper, we firstly make a potential analysis to show the benefit by utilizing the regeneration process, followed by proposing a series of CRDNNs, which combine CNN, RNN, and DNN, to precisely detect Y cycles. Compared to existing algorithms, the CRDNN with bi-directional LSTMs has the best accuracy, and the CRDNN with LSTMs has a comparable performance but much fewer training parameters. Based on our dataset including 119 truck loading cycles, our best neural network shows a 98.2\% test accuracy. Therefore, even with a simple regeneration process, our algorithm can improve the holistic efficiency of mobile machines up to 9\% during Y cycle processes if primary torque concept is used.
\end{abstract}

\begin{IEEEkeywords}
Mobile machines, Y cycle detection, Deep learning, Power management, Hydrostatic drivetrain, Primary torque control
\end{IEEEkeywords}

\section{Introduction}
A new primary torque control concept for hydrostatics mobile machines was introduced in 2018 \cite{rMutschler.2018} and after that, a series of further development has also been made \cite{Xiang.2020}. To date, there are two kinds of mature torque control for hydrostatic drivetrain solution. Unlike the secondary control concept that typically has one or more hydraulic accumulators to build up a constant pressure and controls the output torque by adapting the angle of hydraulic motor, the primary torque control concept controls the pressure in a closed circuit by changing the angle of the hydraulic pump based on a feedback system but without accumulators. Although there are many advantages, primary torque control also has a disadvantage compared to secondary control. While working in Y cycles, which is a typical working process for wheel loaders, secondary control shows an excellent system efficiency even with a simple operation strategy due to its recuperation ability. Without hydraulic accumulators, recuperation is no more possible in a conventional vehicle that has only a combustion engine as its power source, resulting in a lower system efficiency for primary torque controlled mobile machines. Thus, an intelligent operation strategy is needed for primary torque control to improve regeneration performance to compensate for its disadvantage due to the lack of recuperation ability during a Y cycle. Concretely, we use deep learning algorithms to detect Y cycles so that that a current working-process-based operation strategy for primary torque control that can be implemented to improve system efficiency. Apparently, the success of this strategy highly depends on the accuracy of individual working process identification. In this paper, we will focus on the Y cycles detection algorithms. In order to reveal the benefit, we make a potential analysis of regeneration process during Y cycles.

\section{BACKGROUND}

\subsection{Wheel loader}

\begin{figure}[!ht]
\centerline{\includegraphics[width=3in]{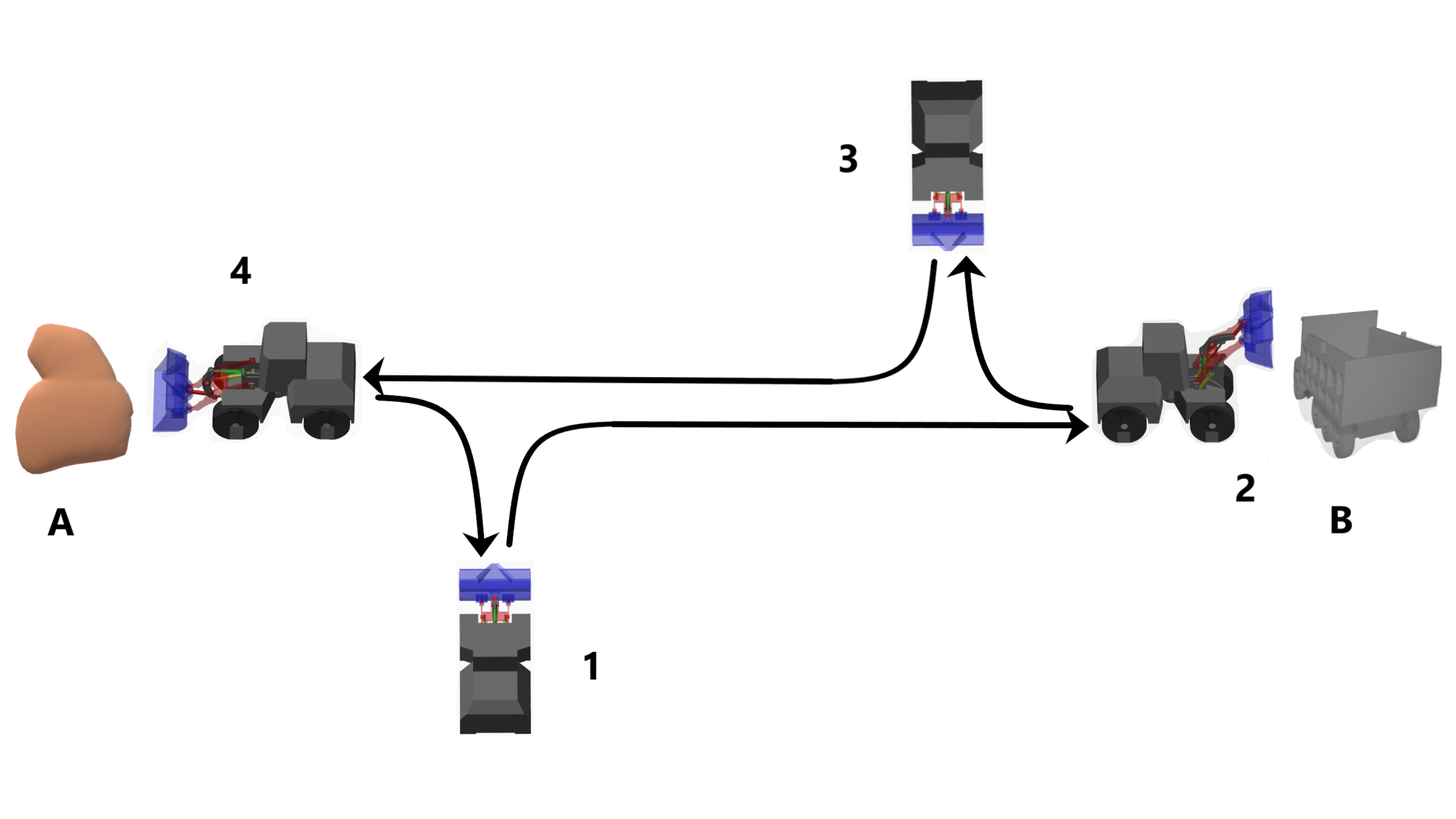}}
\caption{A typical truck loading process (Y cycle)}
\label{fig:fig1}
\end{figure}

The wheel loader is a typical mobile machine used for moving earth. A typical working process is the so-called Y cycle. Concretely, the machine digs the heap and transfers the soil to a truck. During this process, the machine is usually driving in a trajectory similar to a letter Y. Fig. \ref{fig:fig1} illustrates this cycle.

\subsection{Primary torque control}

\begin{figure}[htbp]
\centerline{\includegraphics[width=3.5in]{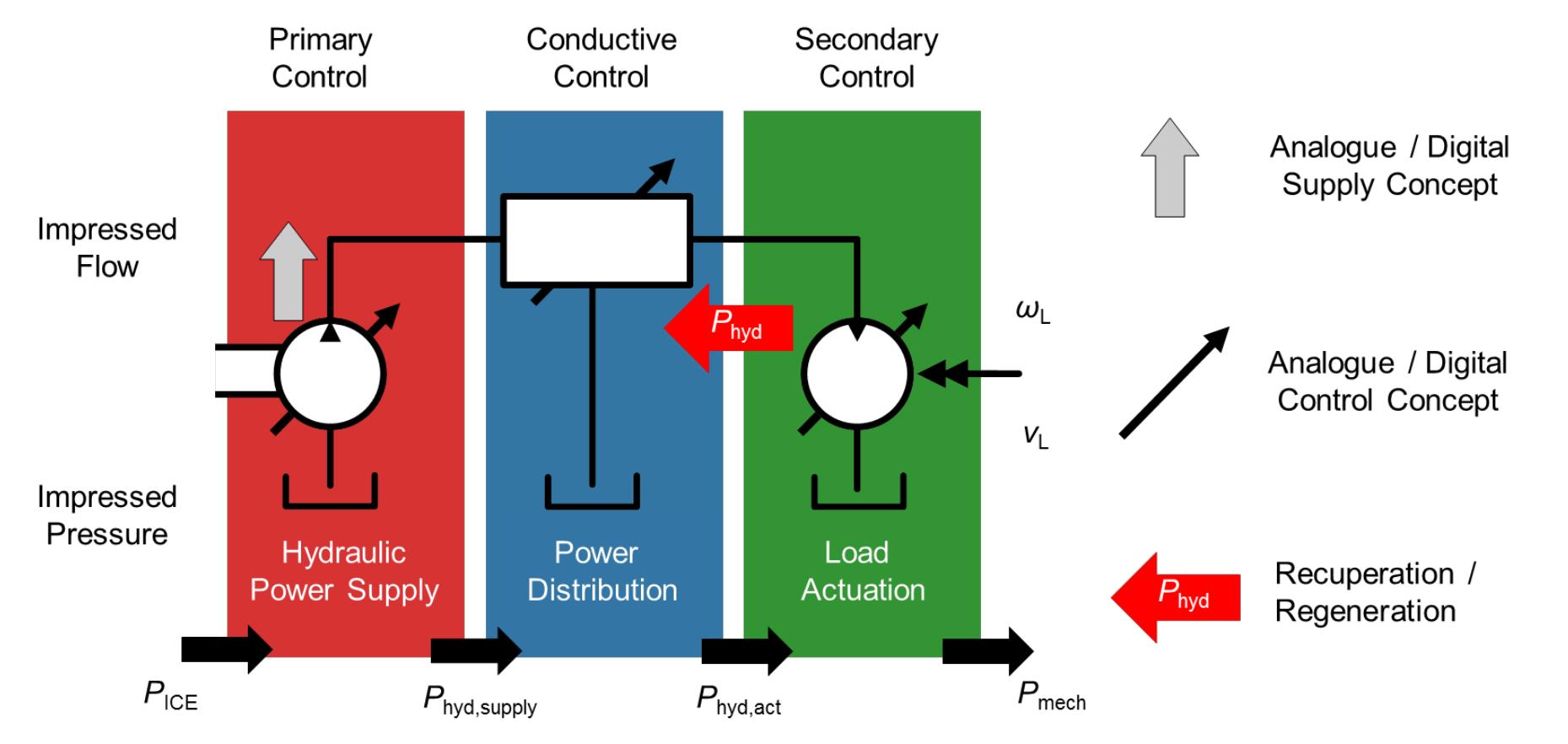}}
\caption{Segment of control concepts \cite{Murrenhoff.2008}}
\label{fig:fig2}
\end{figure}

Murrenhoff has drawn a rule in \cite{Murrenhoff.2008} to classify the different kinds of control concepts, and concrete details are shown in Fig. \ref{fig:fig2}.

One significant advantage of primary torque control is its high efficiency due to the successful introduction of central power management.   The basic idea of central power management derives from the requirement that the power made available to the system should be precisely the same as the power consumed by the system. Besides, in case of power shortage, power management will cut down the power supply to the devices which have a lower priority \cite{rMutschler.2018}.  
To follow this basic idea, every component will compute the energy it requires first, and then the center power manager gathers the information, compares it with the disposable power of power source, and distributes the power to each requester \cite{rMutschler.2018}.

On hydrostatic mobile machines, there is no restrain condition between engine rotation speed and vehicle speed. Thus, optimization of engine efficiency is possible. Generally, the engine speed is set to as low as possible considering the requested vehicle dynamics.

\subsection{Operation strategy}
As aforementioned, a primary torque based wheel loader without a hydraulic accumulator cannot recuperate the energy.

In order to increase the efficiency of our system, we use the regeneration method. Instead of using an additional component, we use our implement to regain the energy. Concretely, we transform the kinetic energy of earthmover to potential energy of the material in the shovel. By the time when the machine decelerates in the truck loading cycle, the hydraulic motor is working as a pump and transfers the energy to lift the shovel.  As a light of that, instead of losing this energy, our operation strategy uses this energy to accelerate the pump for implements on mobile machines.

\subsection{Working process detection algorithms}
Many scholars are interested in utilising machine learning to improve mobile machines regarding efficiency, maintenance, and usability. Especially in the field of working process detection, a series of methods have been drawn.  Pohlandt has used two simple neural networks to predict and recognize the desired work process separately on electrical mobile machines \cite{Pohlandt.2018}. According to his publication, he splits the time series of measured power into many small slip windows to train the simple neural networks. In his research, he found out neural networks might work for some simple cases \cite{Pohlandt.2018}. Another research is from Brinkschulte who points out that the prediction accuracy with bagged trees may dramatically decrease when the drivers have different driving skills \cite{Brinkschulte.2018}. Besides machine learning algorithm, research by Nilsson introduced a method that combined several individually simple techniques including signal processing, state automation techniques, and parameter estimation algorithms. Based on 159 cycles, the accuracy is 93\% \cite{Nilsson.2014}. In 2019, Keller made a case study for an excavator to classify the machine functions using decision tree with an accuracy of 99.97\% without using slip windows \cite{Keller.2019}. In addition, Starke shows that Y cycle can be online recognized with hidden Markov Models (HMM) since HMM was widely used within the context of word recognition to deal with the temporal variability of text or speech \cite{Starke.2019,.}, before 2012 \cite{Hinton.2012}. Also, he pointed out that truck loading is a high variance problem and a simple algorithm should be used owing to the limited of on-board ECU \cite{Starke.2019}.

\section{PROBLEM STATEMENT}
Based on our dataset and previous studies, we summarize the problems faced in this research. 

First of all, the detection of arbitrary Y cycles is a high variance problem. Y cycles are different from site to site. The distance between heap and truck can be quite different. Moreover, drivers are also different. Some drivers have many years of driving experience, and thus have become more aggressive. By contrast, some drivers are still novices who correct themselves during some processes. Last but not least, the materials for transport are different. Therefore, a complicated method is needed to handle the high variance.

Another problem is the limitation of the computing capacity of the ECU on mobile machines. Backpropagation consumes more CPU than forward propagation; thus, online learning usually entails the adoption of a swallow neural networks or an efficiency well-known machine learning algorithm, such as support vector machine (SVM). Consequently, less intelligent learning ability is expected. As a light of that, we dedicate to find an off-line method. Since a simple algorithm might not be really good for dealing with high variance problems, scientists in the fields of Natural language processing (NLP) usually use algorithms that combine many technologies. In the case of HMM, vocal tract length normalization (VTLN) and feature-space maximum likelihood linear regression (fMLLR) are used before HMMs \cite{Hagen.2010}. Neural networks should also be combined together \cite{Sainath.2013}.

\section{WHY WE USE RNN, LSTM}
The initial idea to use long short-term memory (LSTM) is inspired by analogy. Recurrent neural network has been proven to be a powerful tool in the fields of NLP in the past years \cite{Hinton.2012}. One of the significant progress is the introduction of LSTM \cite{Bengio.1994,Hochreiter.1991}. More details about LSTMs can be found in \cite{Hochreiter.1997}. In western countries, clauses are used extensively in writing, making sentences extremely and differently long. Splitting the sentence into many words and using a simple deep neural network (DNN) with a certain number of input layer units, the translation performance is usually unsatisfied. Intuitively, different lengths of sentences make the selection of numbers of input units difficult. A deeper reason lies in the fact that, the simple neural network does not take the sequence of words appearing in the sentence into consideration. With limited input units, simple neural networks can only detect the current situation based on a specific past period. If the decisive information occurred a long time ago, the artificial intelligence (AI) must make its decision based on somewhat useless information causing no wonder a detection mistake. To overcome this problem, LSTM uses update and forget gates to make a shortcut for the vital information to help with the current decision.  Akin to complicated sentences with clause, Y cycles can have very different lengths due to its transport process or workers of different proficiency. As a light of that, LSTM shall solve this principally similar challenge. 

Since our goal is using AI algorithm to detect the working process and thus improve the efficiency of mobile machines by regenerating, “future” information can be used to increase the detection accuracy. Generally, earthmover is first be accelerated in reverse direction and then decelerated after digging into a heap. The duration here implies that even though the algorithm does not recognize directly at the time Y cycle begins, it does not harm the regeneration performance as long as it detects the Y cycle slightly before the deceleration process. Therefore, we also use bi-directional LSTM to improve prediction accuracy.

Similar to HMM that may use some additional technologies to improve its performance, LSTMs also have better performance if convolutional neural networks (CNNs) and DNNs are cooperating together \cite{Sainath.2015,Deng.2014,RazvanPascanu.2014}. The advantages of the combination of CNNs, RNNs, and DNNs, which we call CRDNN in this paper, are shown in the next sections.

\section{POTENTIAL ANALYSIS OF REGENERATION PROCESS}
In this section, a potential analysis is made to show how much energy we can regenerate from the vehicle under the assumption that our detection algorithm works 100\% well, and the regeneration process works perfectly. A study mentioned that vehicle energy is difficult to reuse due to the high rolling friction by mobile machines \cite{Kliffken.2008}. However, according to our observation, most of the experienced drivers flatten the pavement using shovel as a warm-up process. With the preliminary process, they reduce the road roughness to have better driver comfort. Obviously, it also reduces the rolling friction and thereby makes the recuperation or regeneration possible. 
Concretely, we build a wheel loader in Simscape \cite{MATLAB.2019} to simulate Y cycles. The interactions of vehicle mass, vehicle velocity, material mass, and the friction coefficient are studied. Based on our results, the regeneration process can increase system efficiency by about 9\% in general. More details will be found in our other paper. In our simulation, we set the parameters based on the real vehicle and environment data, see Table \ref{tab_1}.


\begin{table}[!ht]
	\centering
	\caption{Parameters of the vehicle and environment}
	\begin{tabular}{l|cl}
	\hline \hline
	Cylinder parameter
		& Value [$m^2$]\\ \hline
	Vehicle mass
		& 10 t \\
	Materials mass
		& 4 t \\
   	Rolling friction coefficient
		& [0.01 0.05 0.3] \\
	Pump loss coefficient
		& 0.8 \\
	Motor loss coefficient
		& 0.8 \\
	Mechanical loss coefficient
		& 0.98 \\

	\hline \hline 
	\end{tabular}
	\label{tab_1}
\end{table}

\section{DEEP LEARNING ALGORITHM}
As aforementioned, we use a series of neural networks to recognize Y cycles. For the sake of simplification, we can say that artificial intelligence is a scientific method for recognizing pattern based on the data with which it has been trained. However, truck loading cycles can be quite different from each other regarding traveling length between heap and truck, driver's skill level, materials, and the dimension of mobile machines. The training of an end-to-end neural network needs a vast dataset, which is still a cost-challenging task today. Instead, we proposed a multi-step approach to detect truck loading cycles. Instead of predicting the Y cycle directly, we firstly predict the loading, the traveling, and the unloading processes since much less data is required for training. Furthermore, after neural networks output its prediction, we might use a modification measure to avoid obvious mistakes. As a light of that, we divide our processes into three sub-processes: vehicle travels, loading, and unloading.

As we mentioned before, we would like to use a more complicated and therefore smarter neural network so that we adopt the off-line learning method to avoid the time-consuming backpropagation.

\subsection{Data acquaint and allocation}
Data is the heart of deep learning.  We split the dataset into training and test data set, 80\% and 20\%, respectively. Besides, we consciously selected three different drivers and did the measurement in different days. Some measurements were done on a rainy day so that the density of material changes. Moreover, we changed the position of heap and truck to vary the length of the Y cycles. The test drivers were not given the information about what we were going to do so that they would behave the same in their daily operations. In short, we consciously increased the diversity of our dataset and tried to include more challenging cases in our dataset. 


The data we fed into the neural networks were selected from the typical sensors on primary torque controlled mobile machines. Concretely, there are the pressure difference inside of the bucket, the vehicle velocity, the vehicle direction signal on joystick, the pressure difference inside of closed-circuit drivetrain, and the pressure difference inside of the boom. The sample rate is 50Hz so that we will not overload ECU.

\begin{figure}[!htbp]
    \centerline{\includegraphics[width=3.5in]{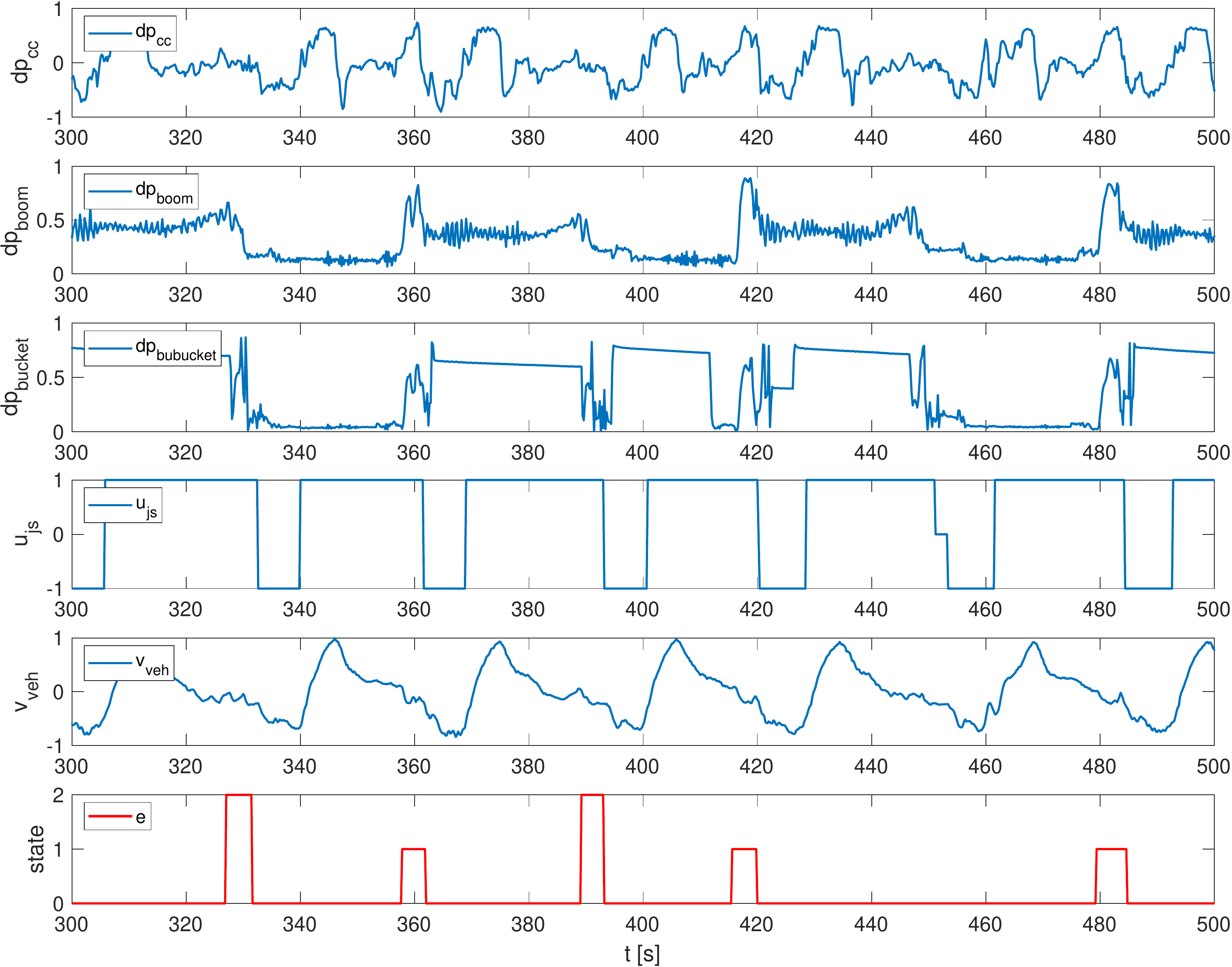}}
    \caption{Normalized measurement data with label}
    \label{fig:fig3}
\end{figure}

Totally, we have created a dataset with 119 Y cycles. 40 of them are gathered from an experienced test engineer, 30 of them are from a development engineer who has aggressive drive behavior, 20 of them are collected when the machine was not well tuned, 29 of them are measured by a senior manager who works many decades in the field of mobile machine. None of the data is collected by a complete layman since we do not think it makes sense. Notice that we have allocated the data collected as the machine with insufficient calibration process into training dataset since it can improve the robustness of our algorithm but not affect our test accuracy. The measurement dataset is labeled as shown in Fig. \ref{fig:fig3}. In the real world, the data is often not perfect. That is, some people may mislabel a tiny portion of data. Therefore, we deliberately labeled some windows as travels through it is actually a loading or unloading process to check the robustness of our algorithms.

Apparently, the variables pressure inside of bucket and vehicle velocity indicates a very strong seasonally. The pressure inside of closed-circuit implies the behaviors of a loading process. Moreover, the signal on joystick demonstrates the state change. Solely based on these variables, an experienced test engineer can tell whether the mobile machine is loading or unloading with almost 100\% accuracy.  Without a doubt, a deep learning model can take over the job to detect Y cycles. However, in the measurement data we have, the Y cycle is not always so regular. For instance, one Y cycle does not always begin with a loading process, followed by an unloading process. The driver might think he has loaded too small amount of load so he comes back after a small reversing process and digs into the heap again. It happens when the driver is not so skilled or is mistaken. Such a case increases the difficulties of detecting the truck loading process by deep learning.

\subsection{Data preparation}
The measurement data were not pre-treated by human observation depending on the dataset before fed into neural networks even we agree a pre-processing can surely increase the accuracy of prediction. The reason for that is we are worried about the pre-processing may exaggerate the performance of neural network since some pre-processing are almost impossible in reality. Therefore, we did use the non-adaptive method to prepare the dataset: only a first-order system is used to smooth our data. After that, the dataset will be split into small slip windows. If the size of time windows is 10 sample times, the events in the past 2 seconds are taken into consideration since our sample frequency is 50Hz. Slip windows are used for avoiding the influence of data too long ago. 

The measured data is normalized before training since we want to avoid one single variable that has too much influence on each gradient descent step. As a result of that, the cost function's shape changes into a more spherical one rather than a high curvature ellipse one. 

Besides, the labeled date is converted into one hot vector to have the same categorical value, as shown in eq. \eqref{eq1},
\begin{equation}
    Y^{(1)} = \left[ \begin{array}{ccc}
        0 & 1 & 0
    \end{array}\right]
    \label{eq1}
\end{equation}

which demonstrates that the 1st sample is labelled as loading.
In our dataset, 11.62\% of all working time is in the loading process, and 7.86\% is in the unloading process. Obviously, our dataset for truck loading process is skewed. That means, even if we always predict that we are neither in loading nor unloading process, we have a test accuracy at about 80\%. To avoid it, we should use confusion matrices and micro average F1 scores to evaluate the performance of our algorithm. Based on an exploratory training, the training cost without anti overfitting goes down to an extremely low level while the test cost goes firstly down and then explodes up. This indicated that our dataset has well considered the variance of Y cycle in different cases.

\section{COMBINED NEURAL NETWORKS}

\begin{figure*}[!htbp]
    \centerline{\includegraphics[width=7in]{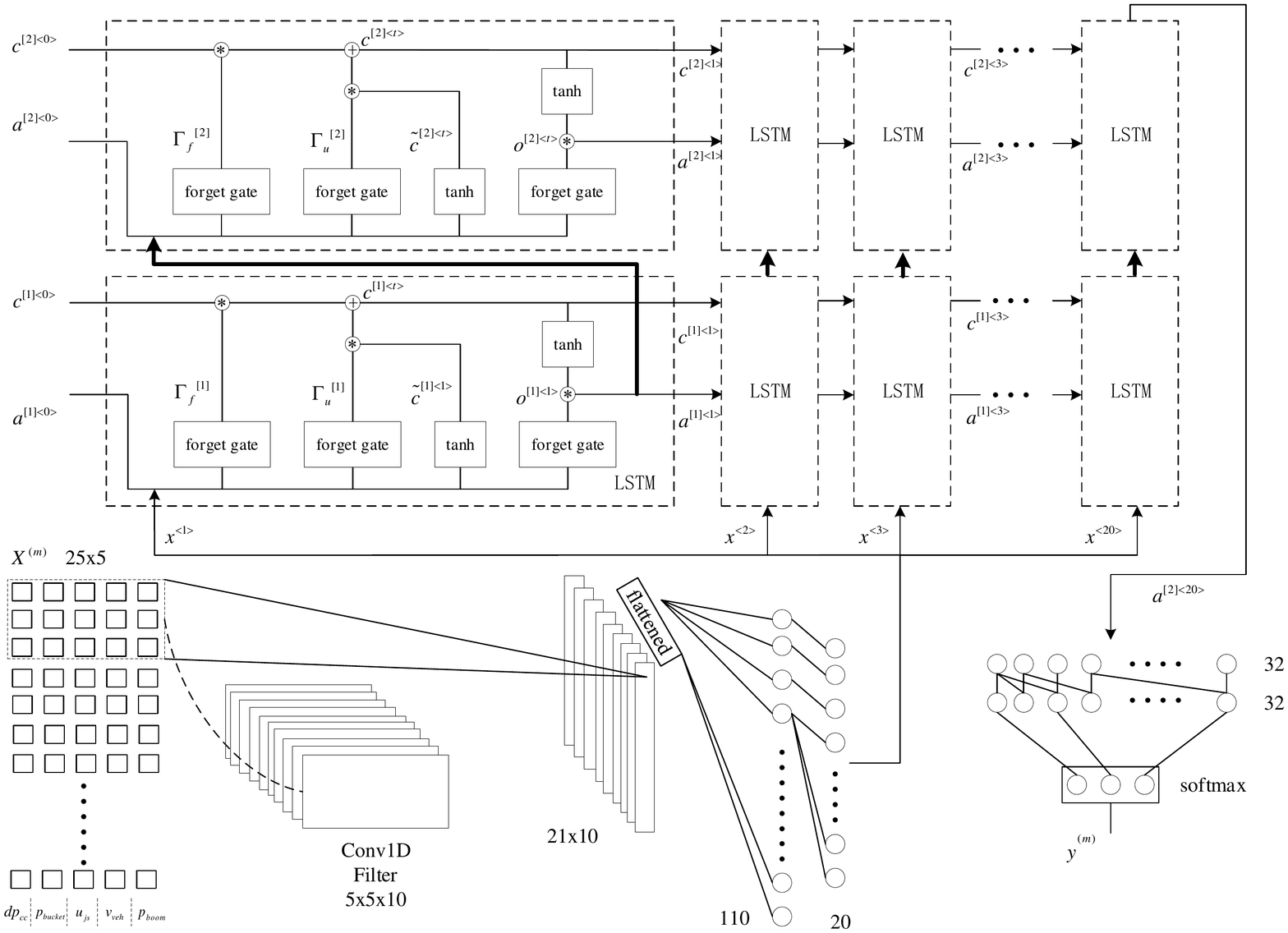}}
    \caption{Detailed description of CRDNN with two LSTMs}
    \label{fig:fig4}
\end{figure*}

To have a better detection performance, we take the advantages of combined neural networks, CRDNNs. We are going to explore the combination of CNNs, RNNs, and DNNs in this section. They all have limitations so that we believe that combined neural networks can be complementary for the disadvantages of each other. For example, LSTMs are good at temporal modelling while they cannot have a more significant number of hidden layers.

As shown in Fig. \ref{fig:fig4}, we use one onedimensional onvolutional neural network (conv1D) at the beginning to provide better features for LSTMs. It is followed by two DNNs to reduce the dimension of the output of CNN. Further, we add two LSTMs since it is considered as an excellent tool for many time series applications. At the end, two DNNs are used to increase nonlinear hidden layers and thus increase the prediction performance by making a deeper mapping. The core of LSTMs is the update and forget gate to handle the long and short term data. Eq. \eqref{eq2} demonstrates the idea.
\begin{equation}
    \left\{ \begin{array}{c}
        \tilde{c}^{\langle t \rangle} = \tanh{(W_c[a^{\langle t-1 \rangle}, x^{\langle t \rangle}] + b_c)}\\
        \Gamma_u = \sigma(W_u[a^{\langle t-1 \rangle}, x^{\langle t \rangle}] + b_u)\\
        \Gamma_f = \sigma(W_f[a^{\langle t-1 \rangle}, x^{\langle t \rangle}] + b_f)\\
        \Gamma_o = \sigma(W_o[a^{\langle t-1 \rangle}, x^{\langle t \rangle}] + b_o)\\
        {c}^{\langle t \rangle} = \Gamma_u*\tilde{c}^{\langle t \rangle}+\Gamma_f*{c}^{\langle t-1 \rangle}\\
        a^{\langle t \rangle} = \Gamma_o*\tanh{{c}^{\langle t \rangle}}
        \end{array} \right.
    \label{eq2}
\end{equation}

Generally, the learning ability is increasing as the number of hidden layers increase. However, more hidden layers result in much more training parameters that may be a heavy load for vehicle ECU. In this section, we evaluate the CRDNNs' test accuracy regarding the hidden layers, the units in a hidden layer, and time windows. \\
Theoretically, LSTMs can work without a slip window. However, we need to avoid the data before a disruptive event, such that the driver stops the vehicle to relax for a while, which affects the prediction performance. Therefore, we also use the slip windows for CRDNNs. The window size can affect the performance of neural networks since a larger window size allows the neural networks considered a more extended period to make the decision. 

\begin{figure*}[!t]
        \newcommand{\w}{0.25}
        \centering 
        \subfloat[1 LSTM (ws = 9)]{
            \includegraphics[width=\w\textwidth]{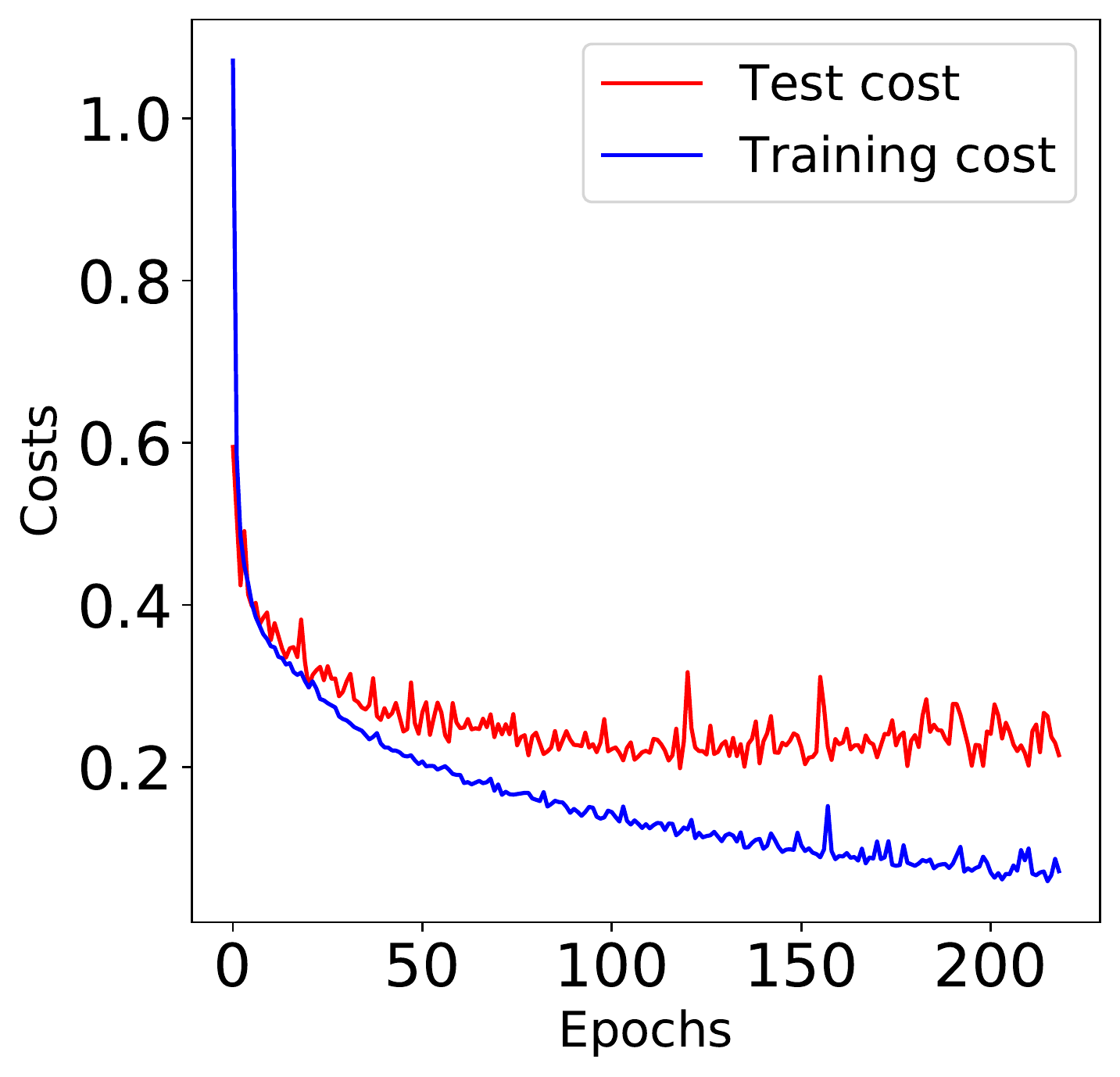}}
            \hfil 
        \subfloat[2 LSTMs (ws = 9)]{
            \includegraphics[width=\w\textwidth]{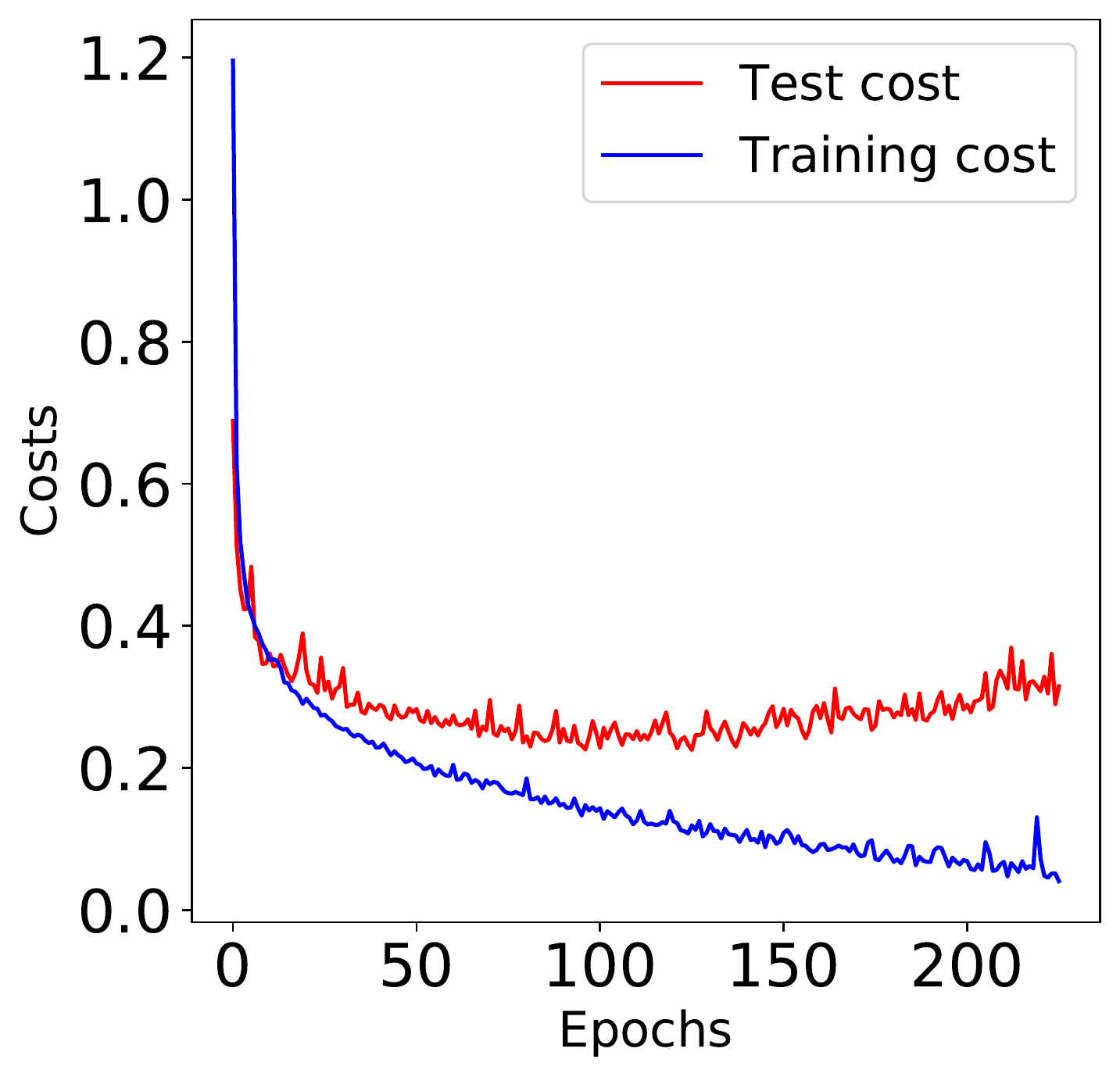}} 
            \hfil
        \subfloat[2 BiLSTMs (ws = 9)]{
            \includegraphics[width=\w\textwidth]{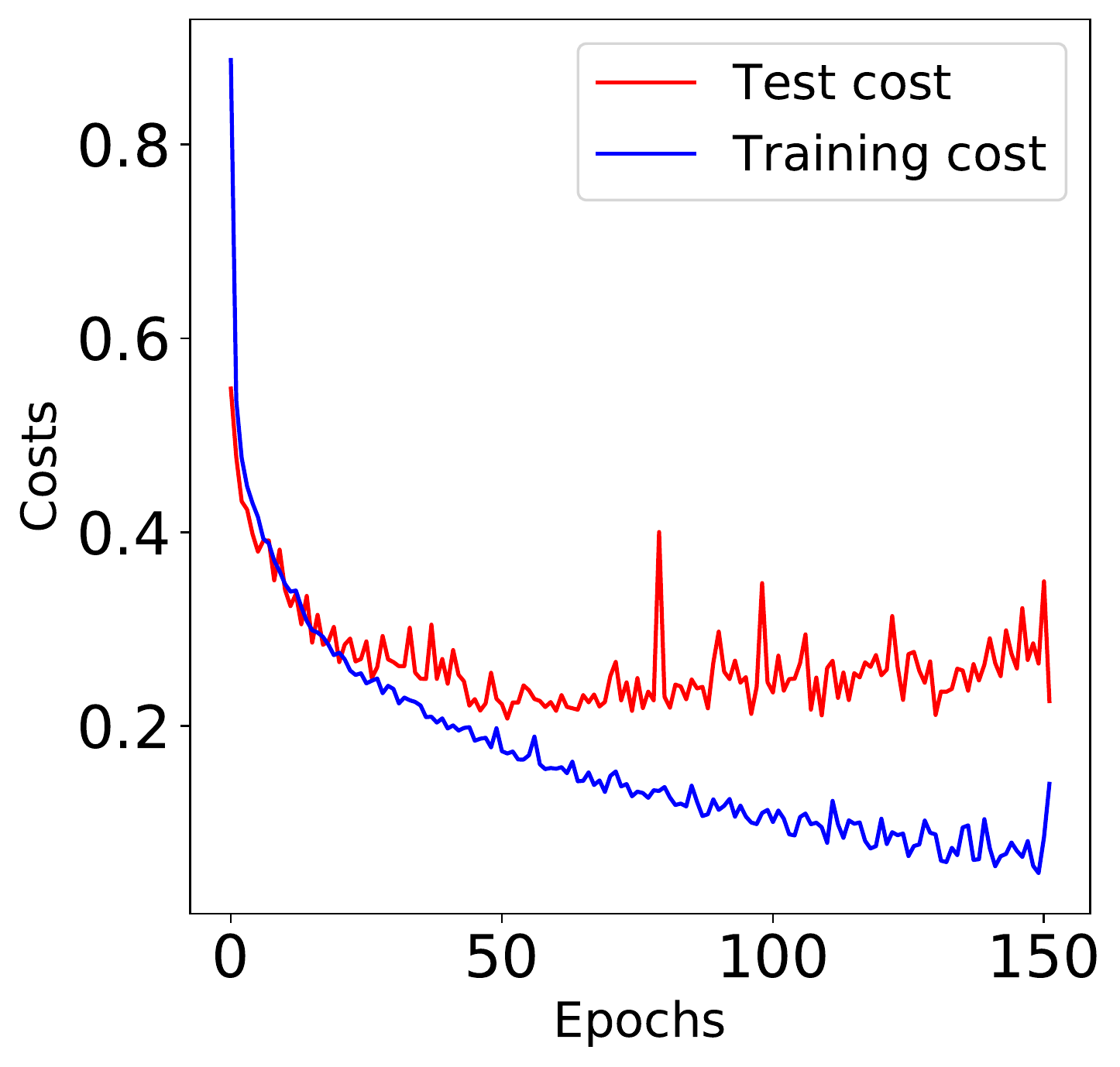}} 
            \hfil
        \subfloat[1 LSTM (ws = 15)]{
            \includegraphics[width=\w\textwidth]{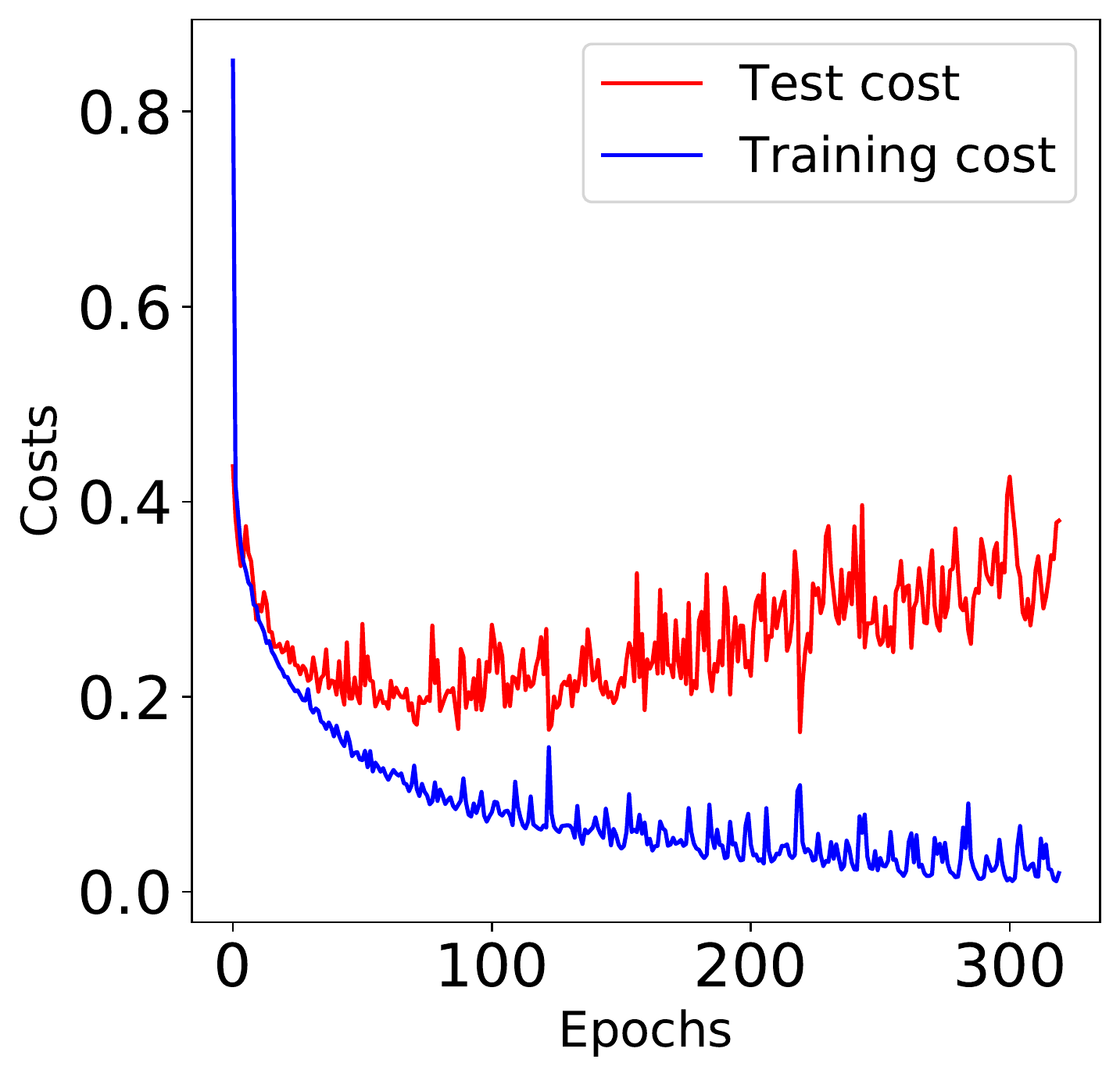}}
            \hfil 
        \subfloat[2 LSTMs (ws = 15)]{
            \includegraphics[width=\w\textwidth]{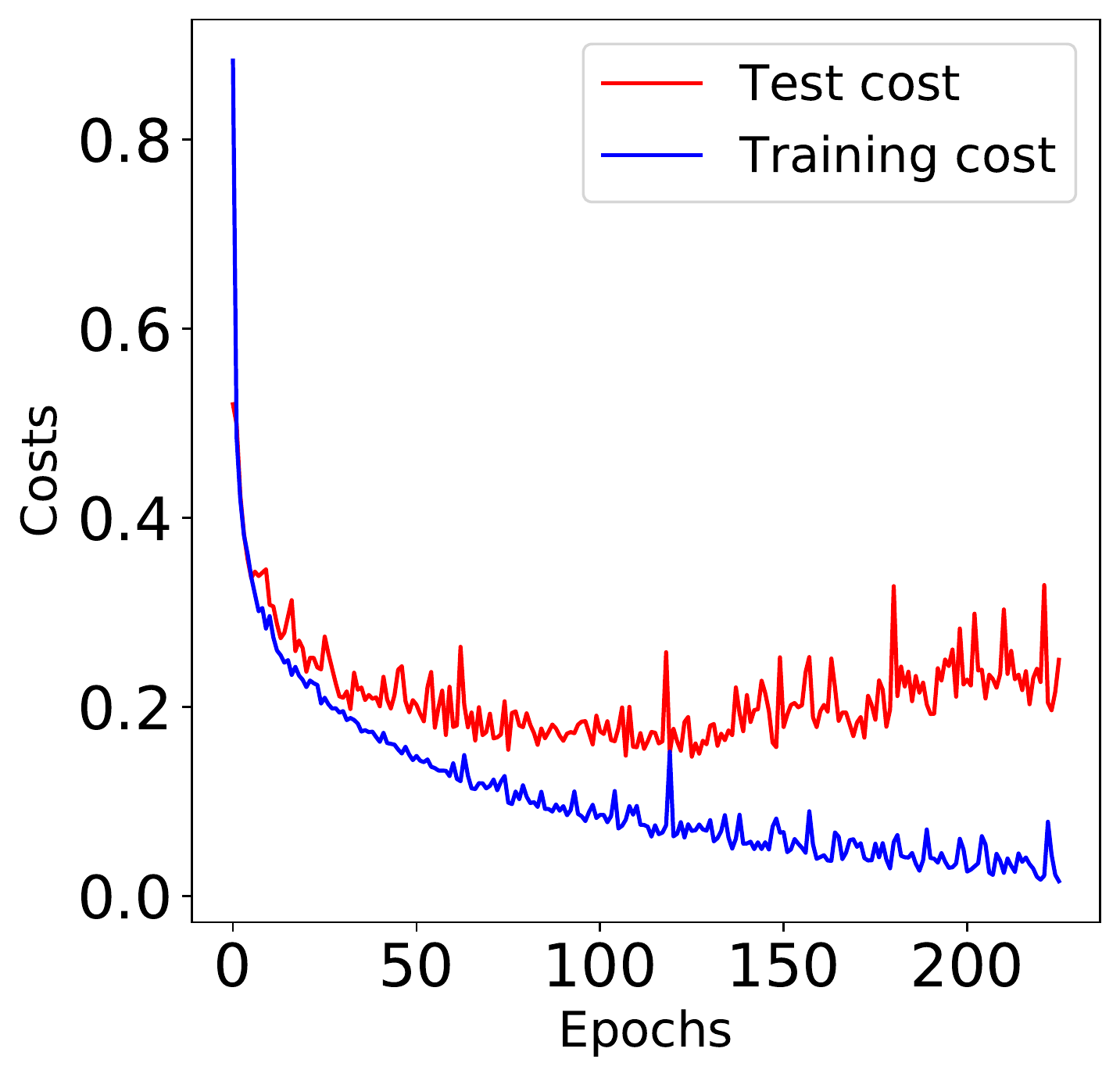}} 
            \hfil
        \subfloat[2 BiLSTMs (ws = 15)]{
            \includegraphics[width=\w\textwidth]{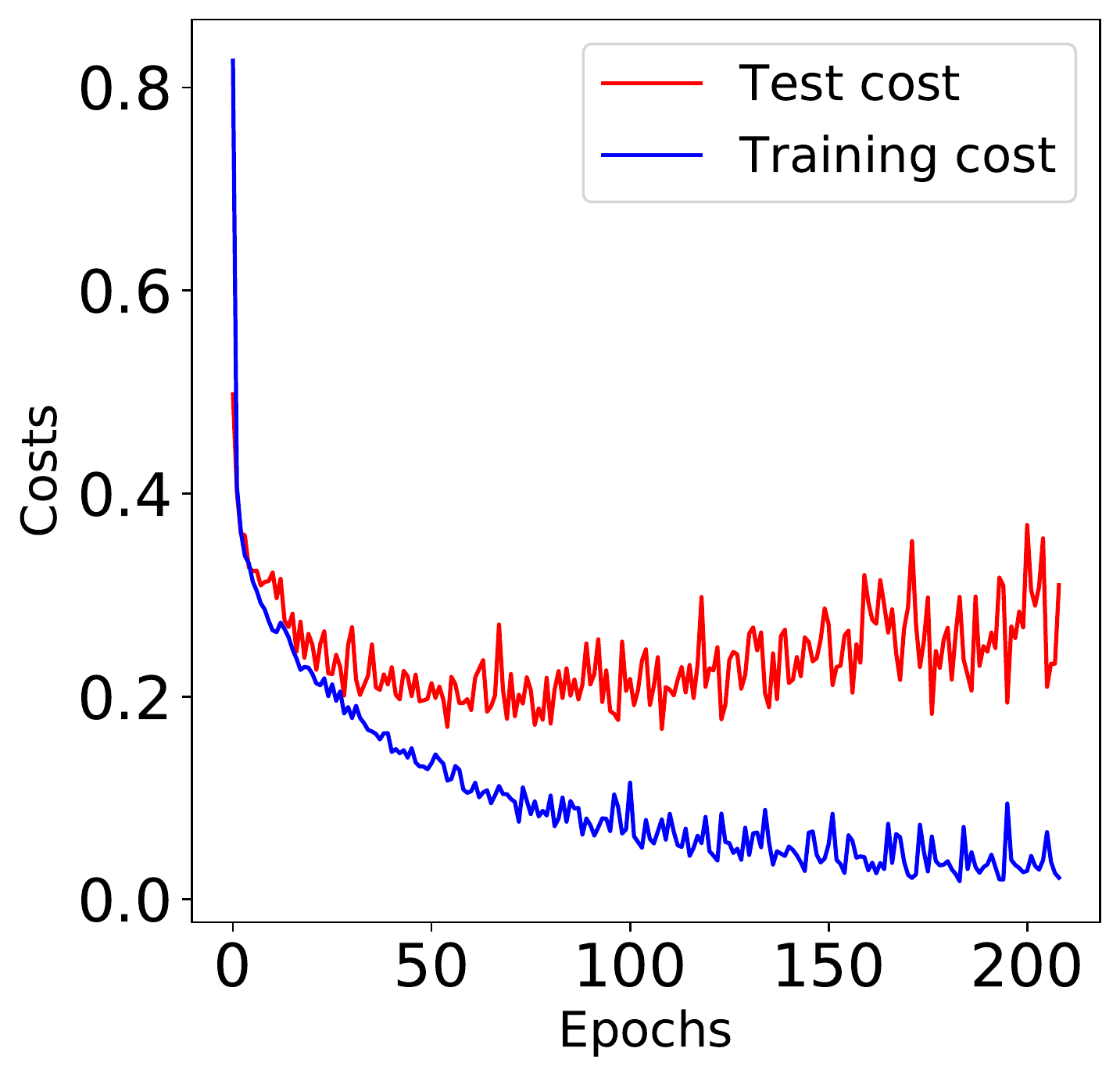}} 
            \hfil
            \subfloat[1 LSTM (25)]{
            \includegraphics[width=\w\textwidth]{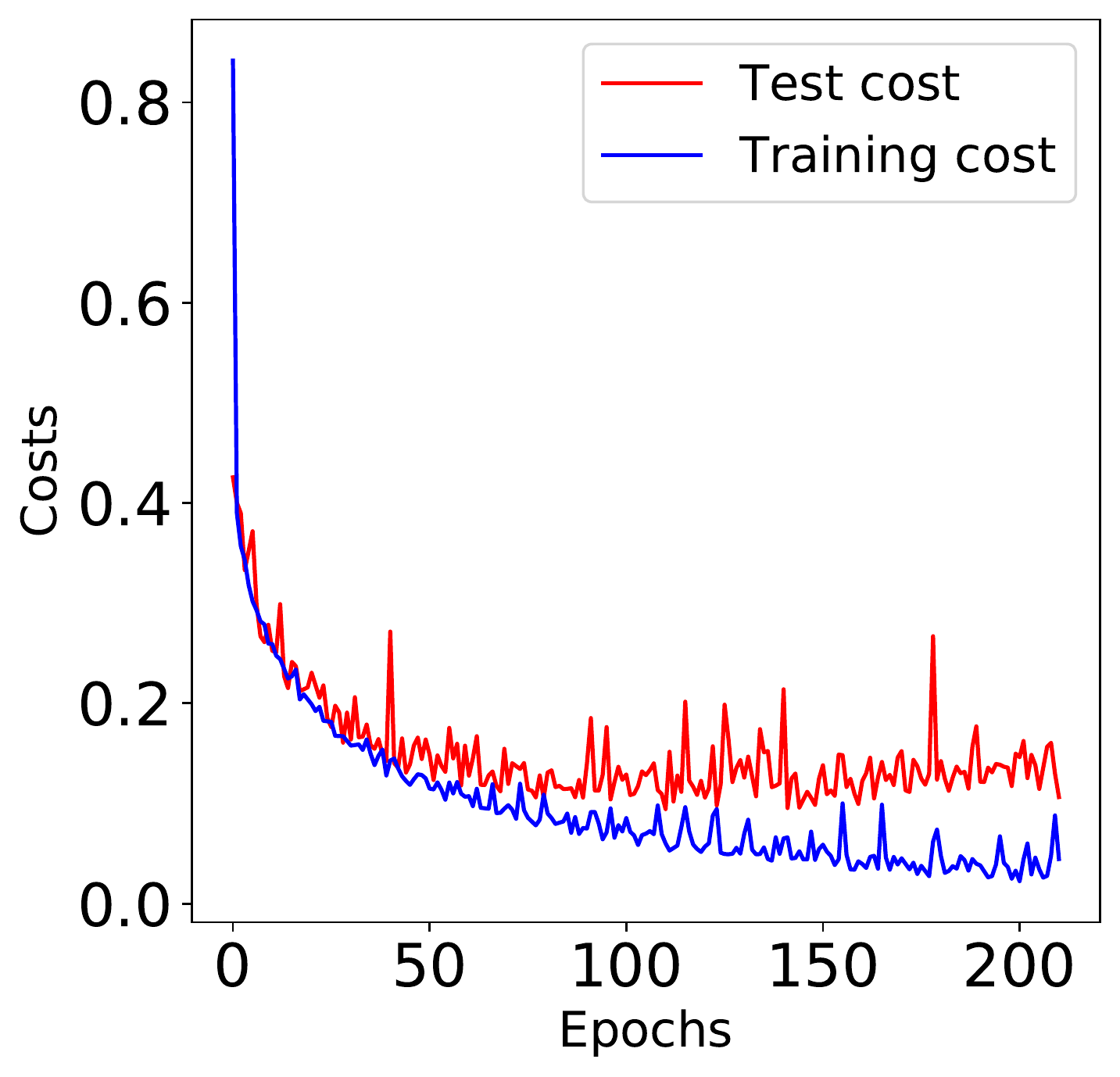}}
            \hfil 
        \subfloat[2 LSTMs (ws = 25)]{
            \includegraphics[width=\w\textwidth]{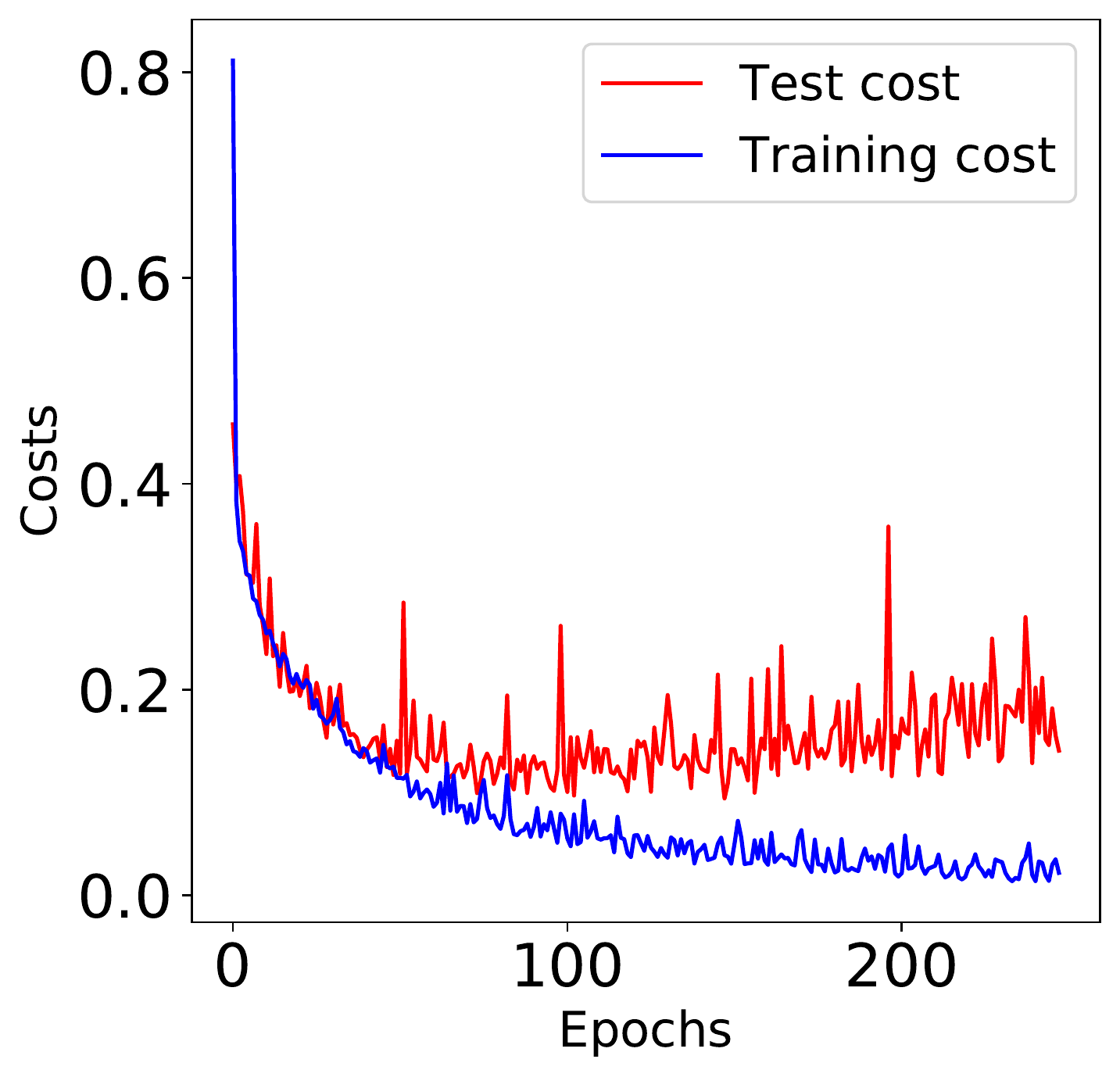}} 
            \hfil
        \subfloat[2 BiLSTMs (ws = 25)]{
            \includegraphics[width=\w\textwidth]{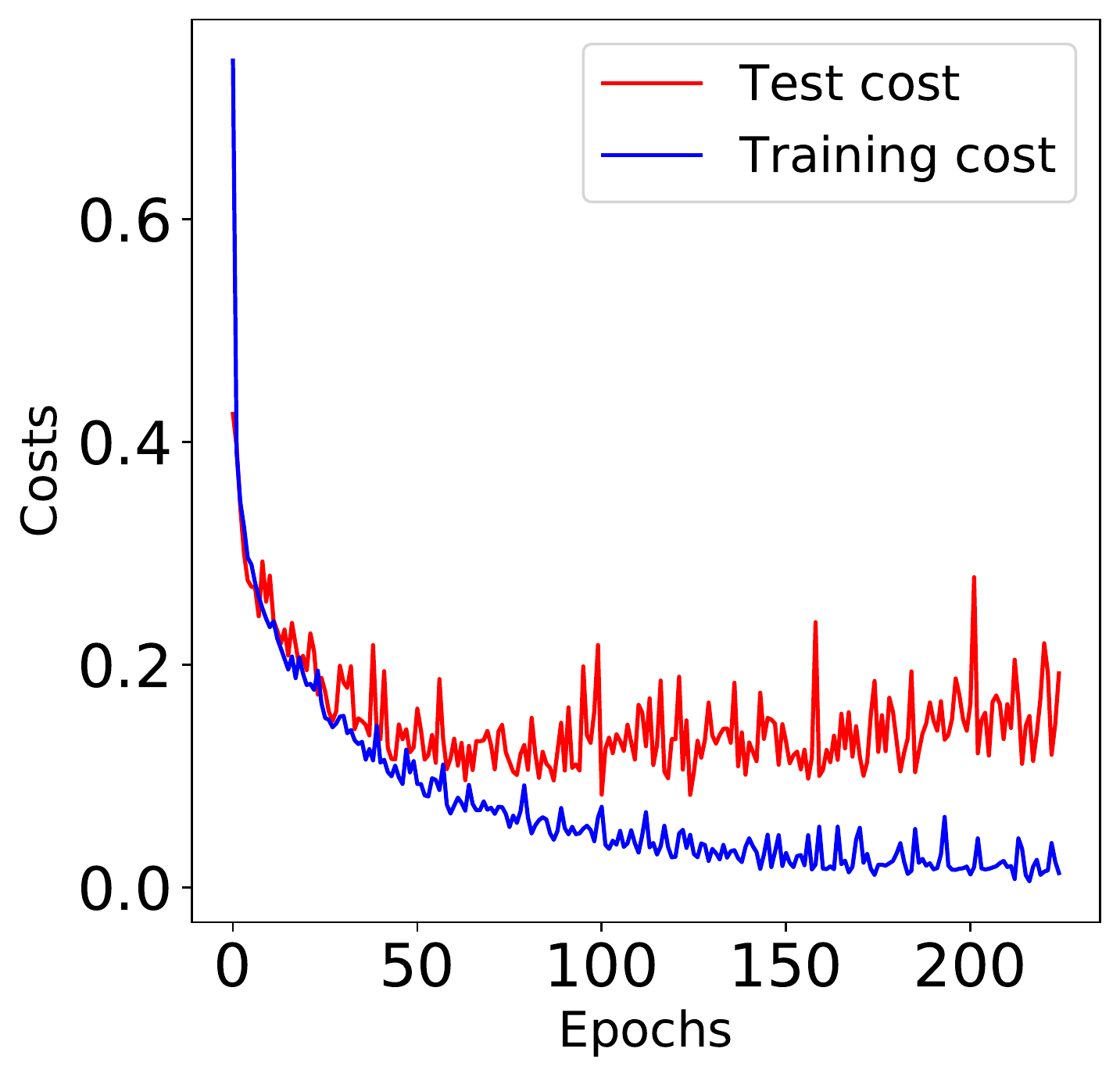}} 
            \hfil
        \caption{Training- and test costs versus epochs} \label{fig:fig5}
\end{figure*}

Since we want to know which model has the best test accuracy and which one has a good test accuracy but with fewer training parameters, we show the different performances of different architecture with different window sizes. We supervise the training- and test costs over epochs and stop the optimization process when there is a noticeable tendency that test cost increases. The training cost versus iteration of different neural networks is shown in Fig. \ref{fig:fig5}. For example, in the case of CRDNN with 2 LSTMs that is fed the data with a window size of 9, we stop the iteration at epoch 60.

To find out the suitable hyper-parameters of neural networks, we analyze the weights of each layers of neural networks. However, while people recognize the working process mainly by watching the pressure inside of bucket, CRDNNs do not pay too much attention to this variable since the absolute value of weight for it is no considerably larger than the others.

\begin{figure*}[!t]
        \newcommand{\w}{0.25}
        \centering 
        \subfloat[1 LSTM (ws = 9)]{
            \includegraphics[width=\w\textwidth]{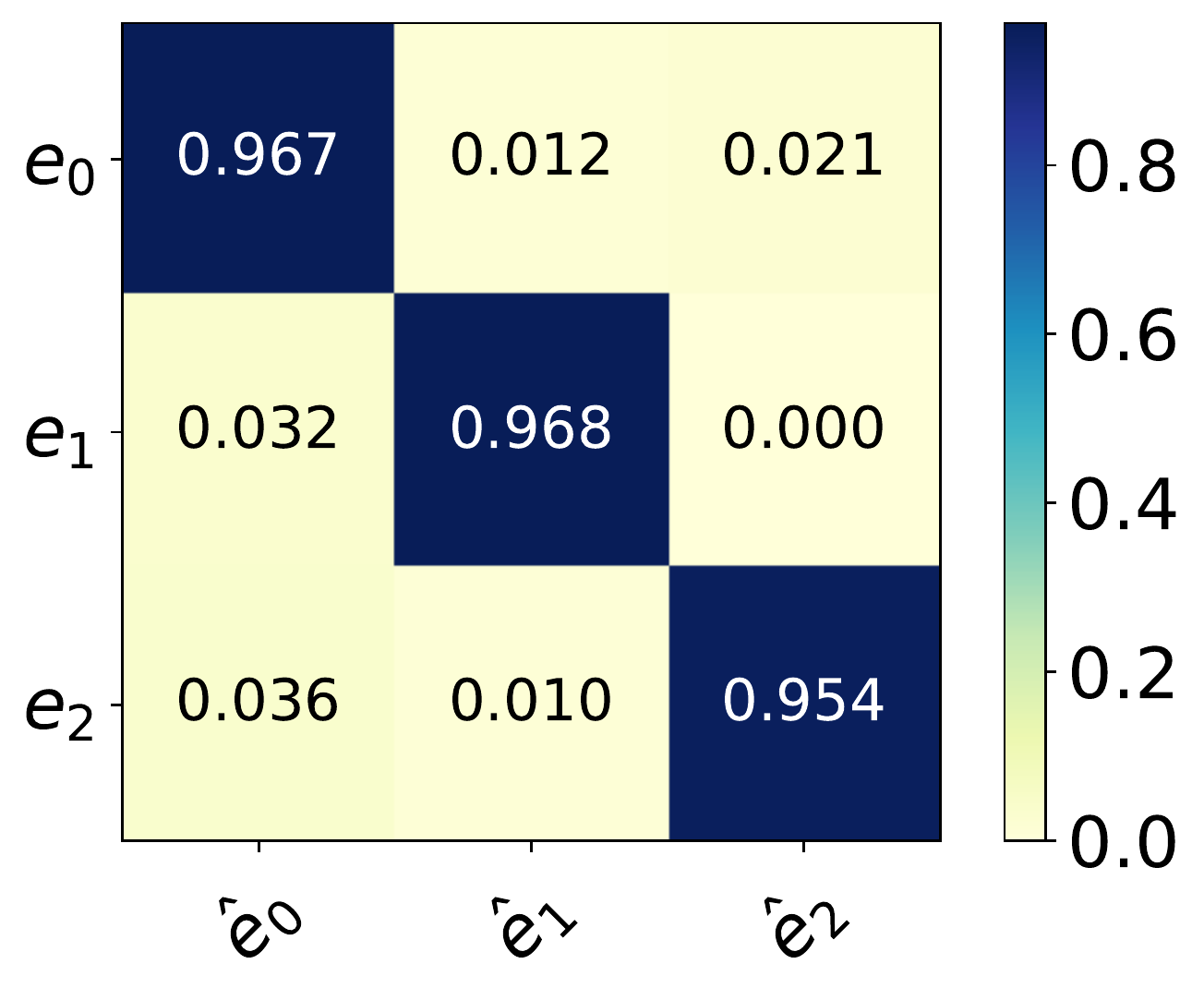}}
            \hfil 
        \subfloat[2 LSTMs (ws = 9)]{
            \includegraphics[width=\w\textwidth]{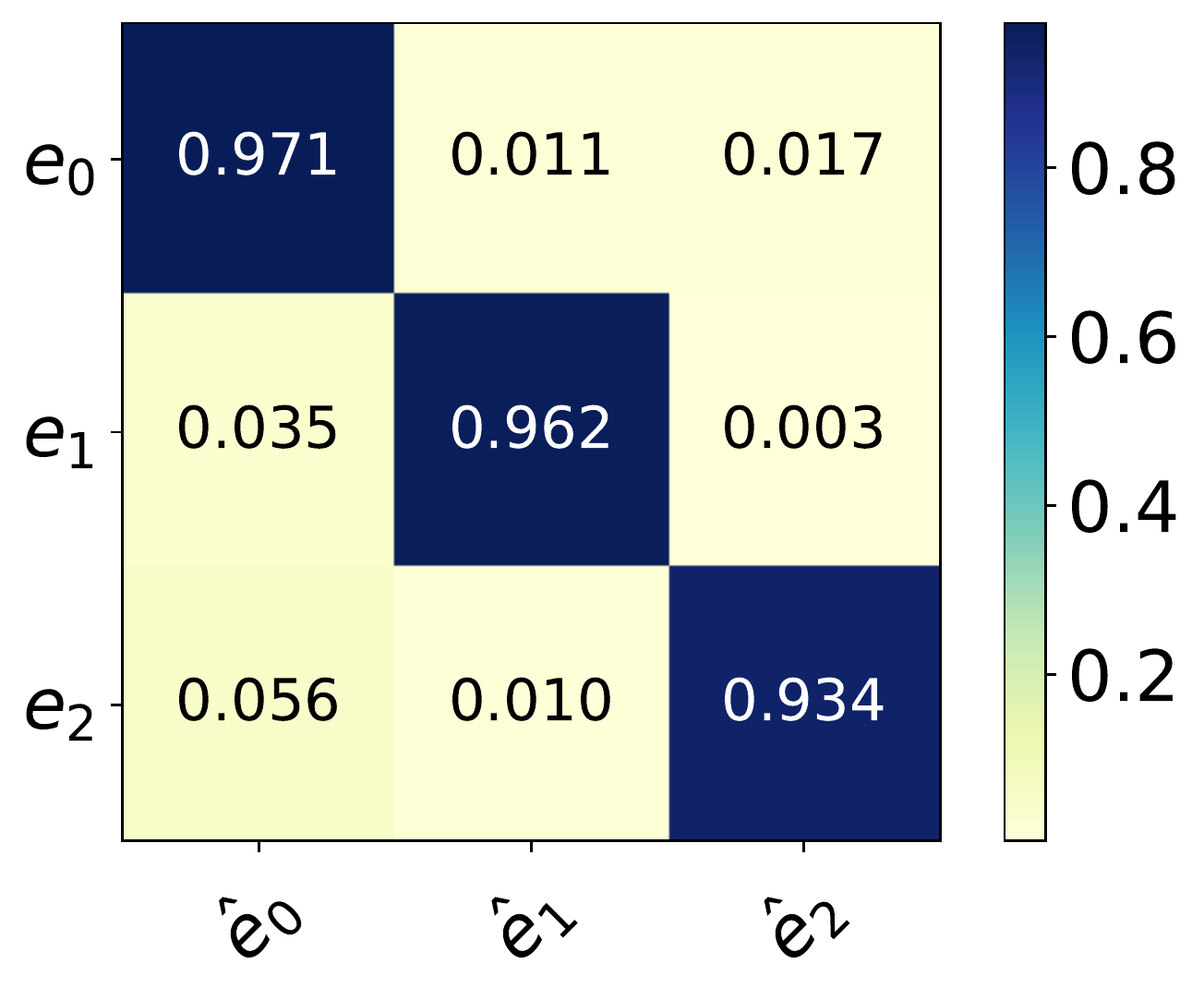}} 
            \hfil
        \subfloat[2 BiLSTMs (ws = 9)]{
            \includegraphics[width=\w\textwidth]{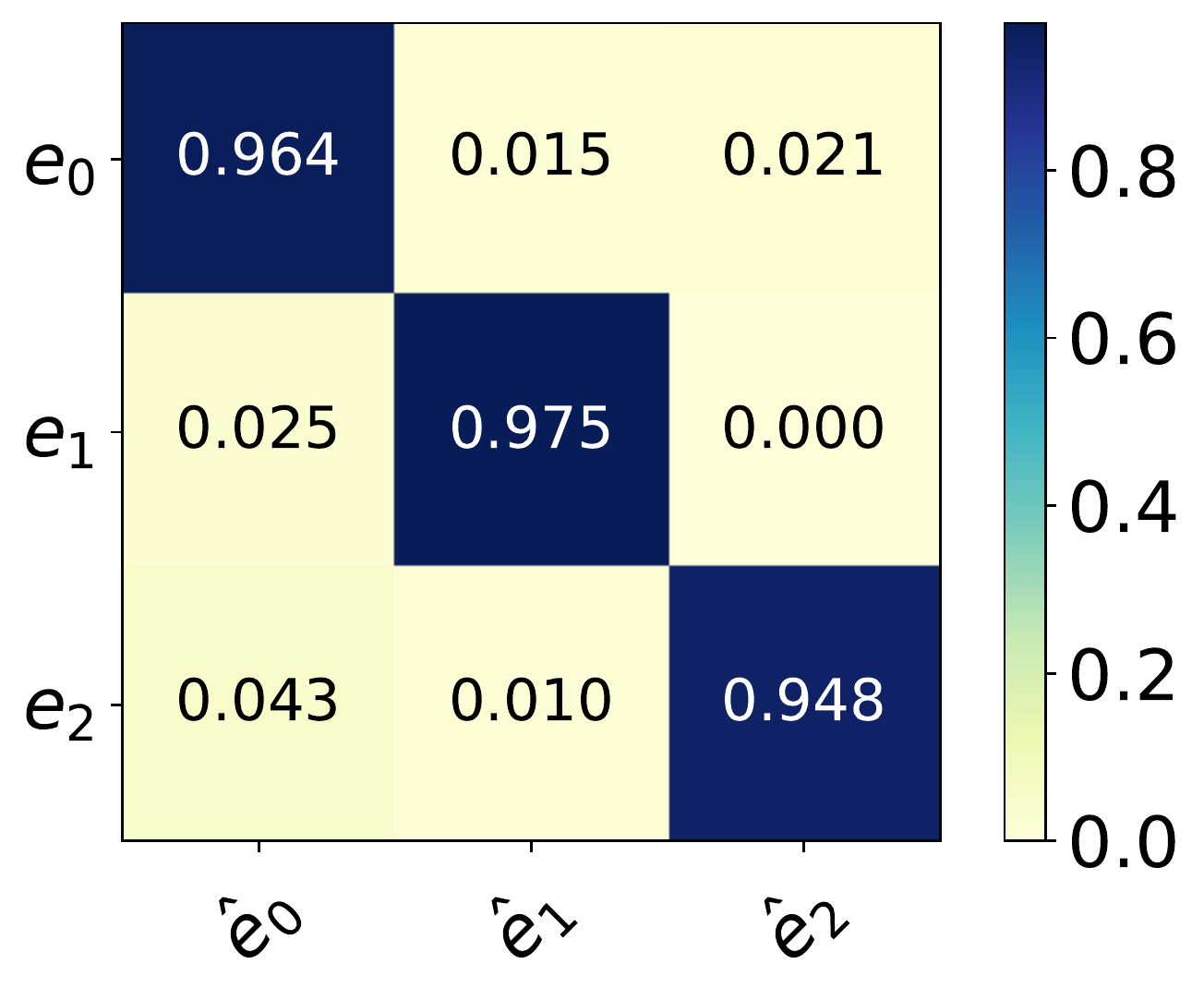}} 
            \hfil
        \subfloat[1 LSTM (ws = 15)]{
            \includegraphics[width=\w\textwidth]{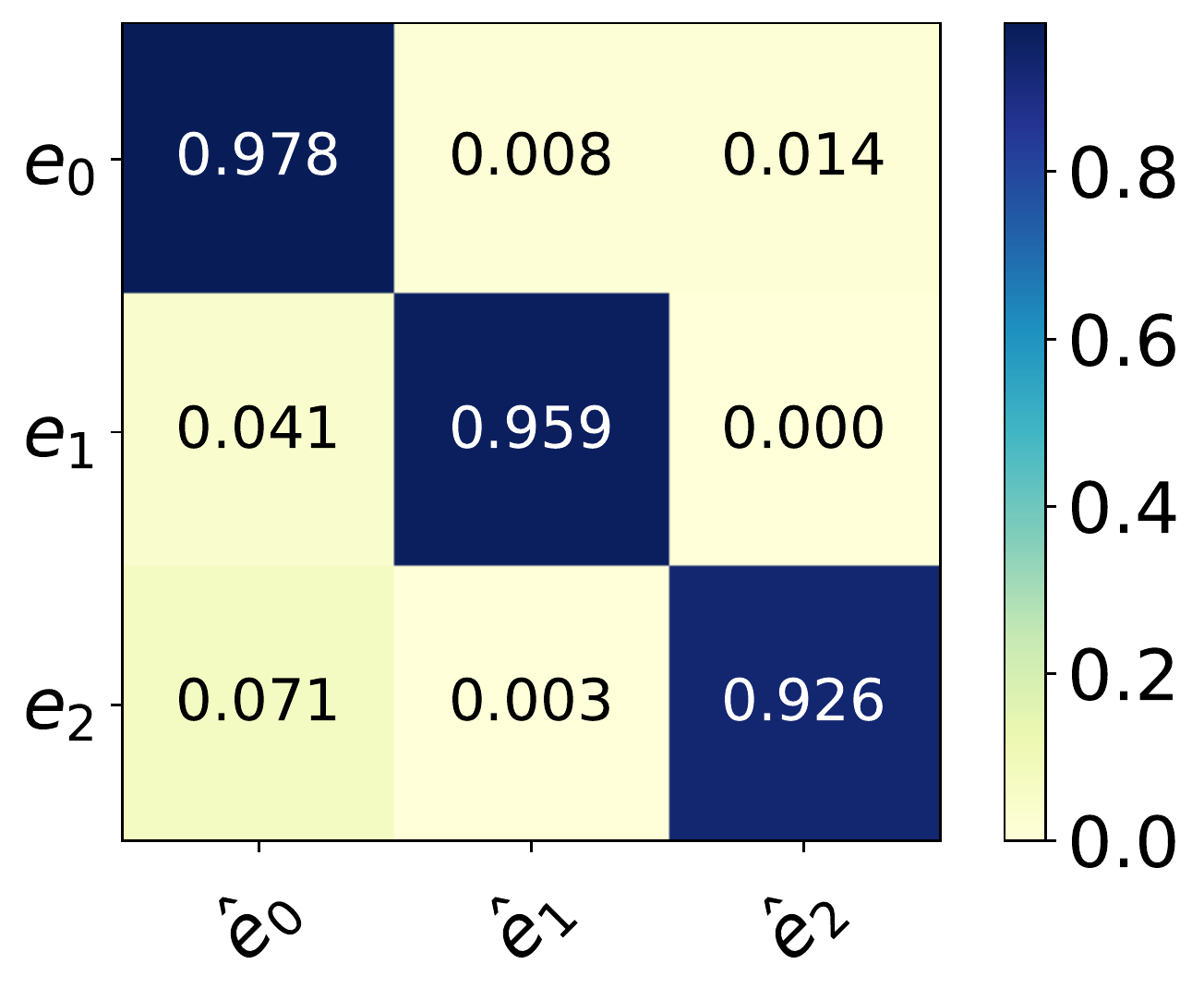}}
            \hfil 
        \subfloat[2 LSTMs (ws = 15)]{
            \includegraphics[width=\w\textwidth]{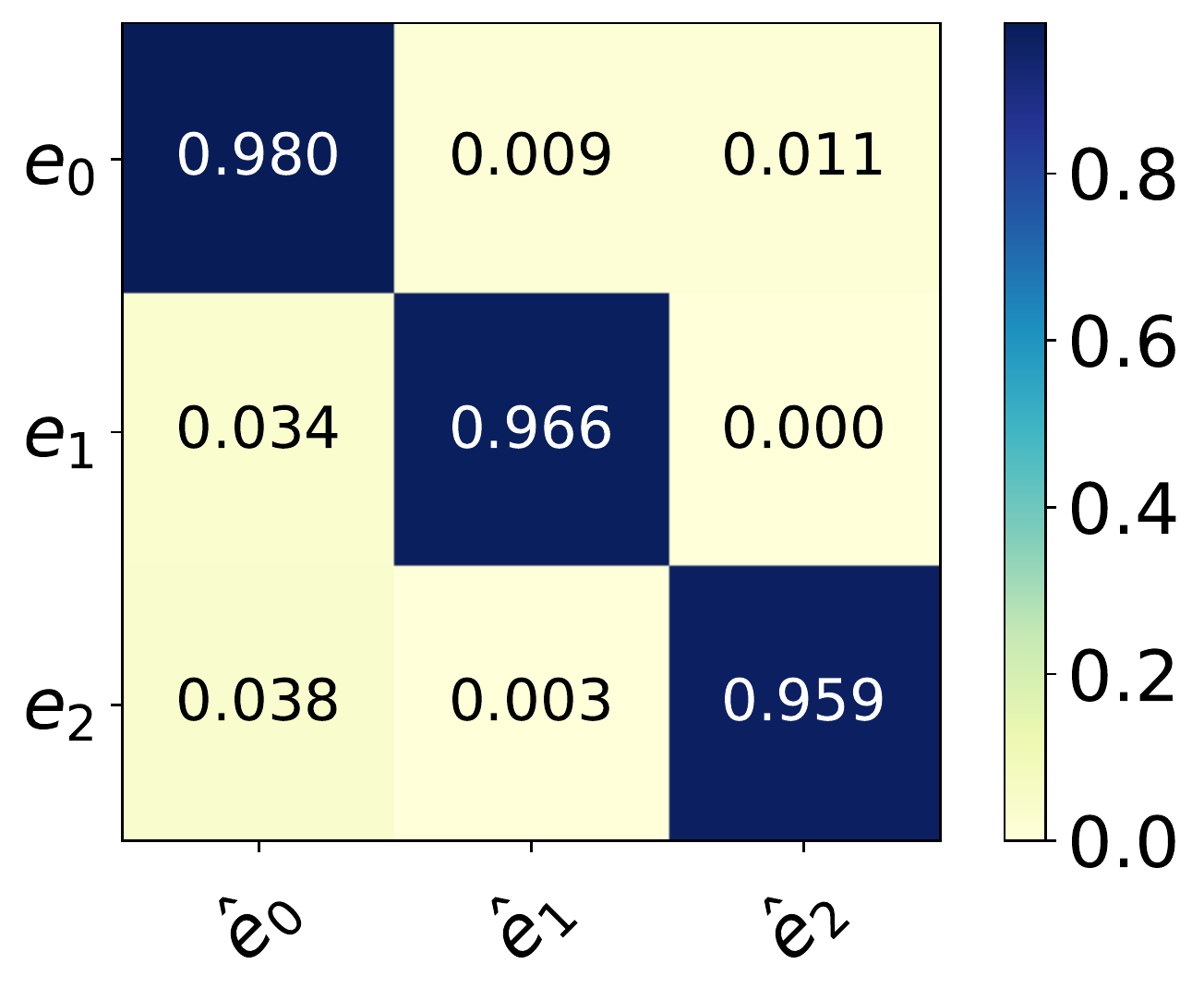}} 
            \hfil
        \subfloat[2 BiLSTMs (ws = 15)]{
            \includegraphics[width=\w\textwidth]{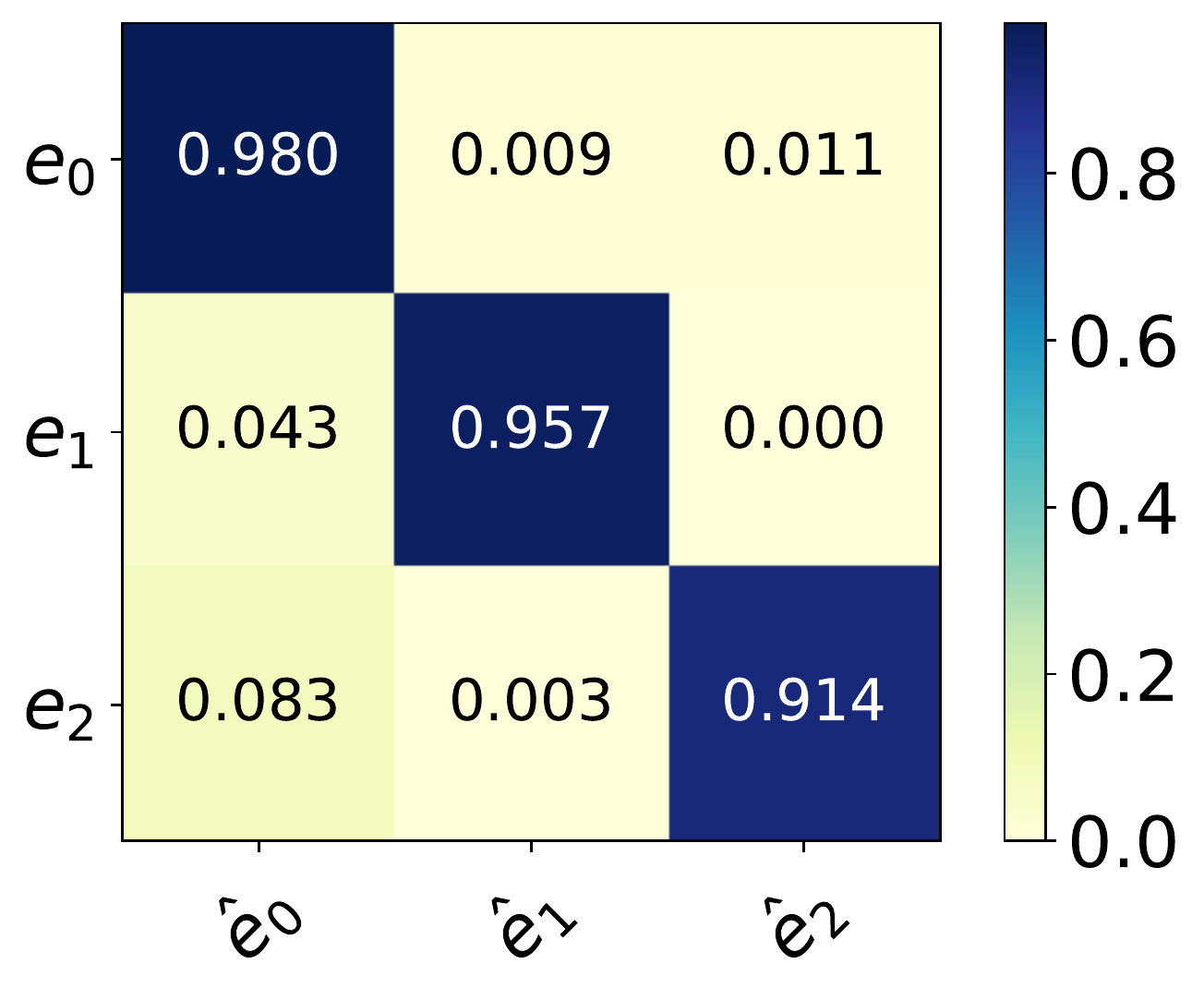}} 
            \hfil
            \subfloat[1 LSTM (25)]{
            \includegraphics[width=\w\textwidth]{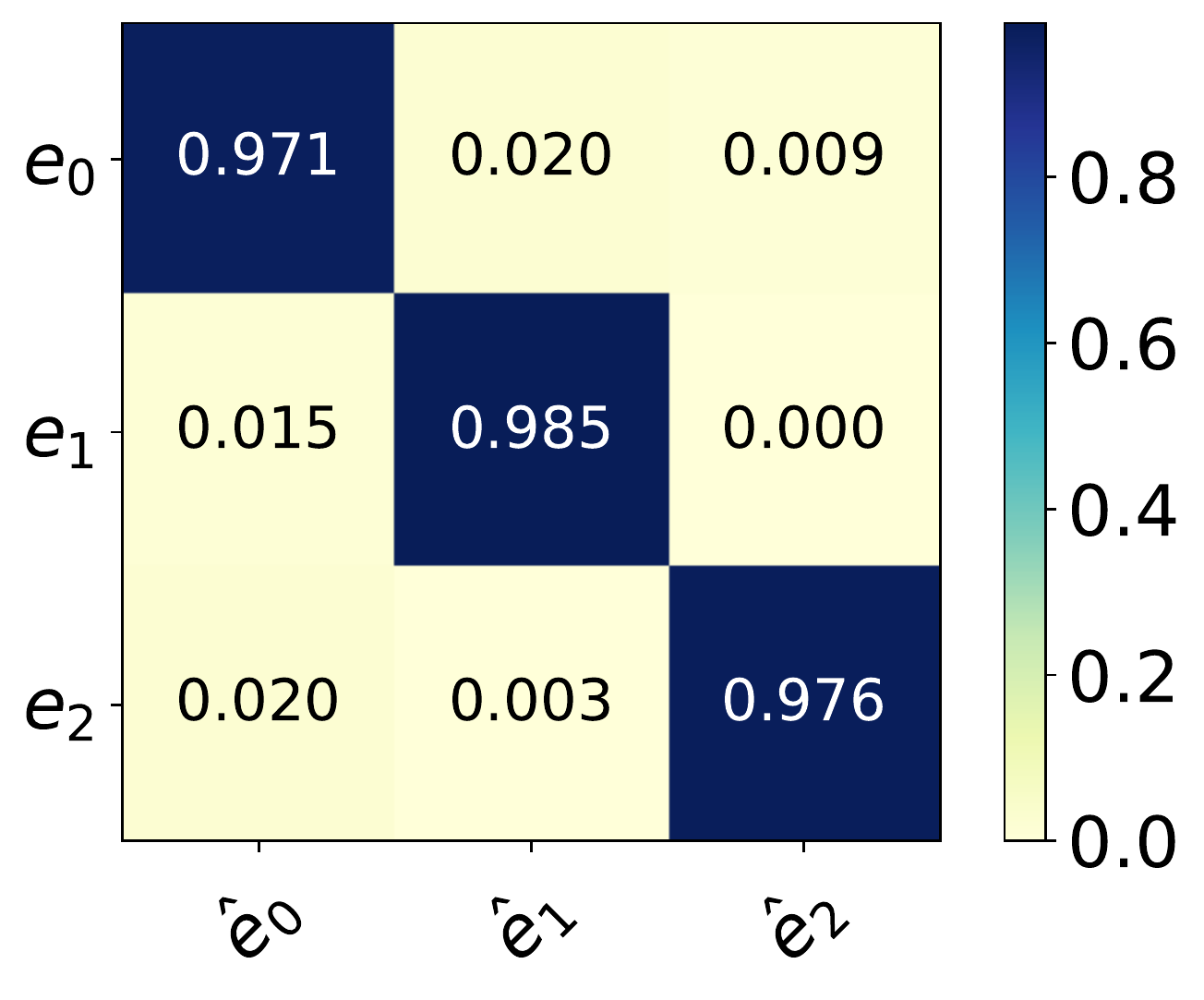}}
            \hfil 
        \subfloat[2 LSTMs (ws = 25)]{
            \includegraphics[width=\w\textwidth]{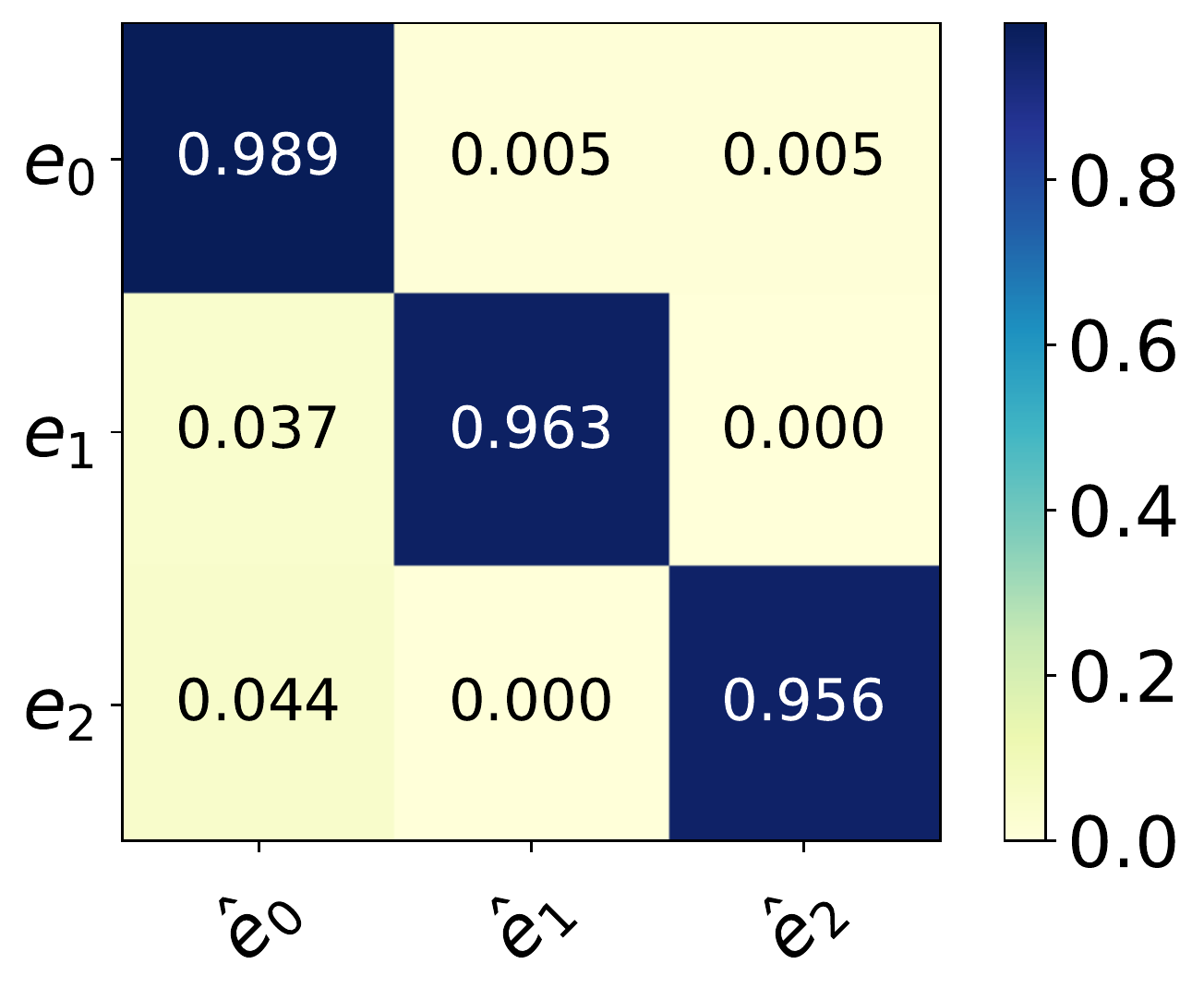}} 
            \hfil
        \subfloat[2 BiLSTMs (ws = 25)]{
            \includegraphics[width=\w\textwidth]{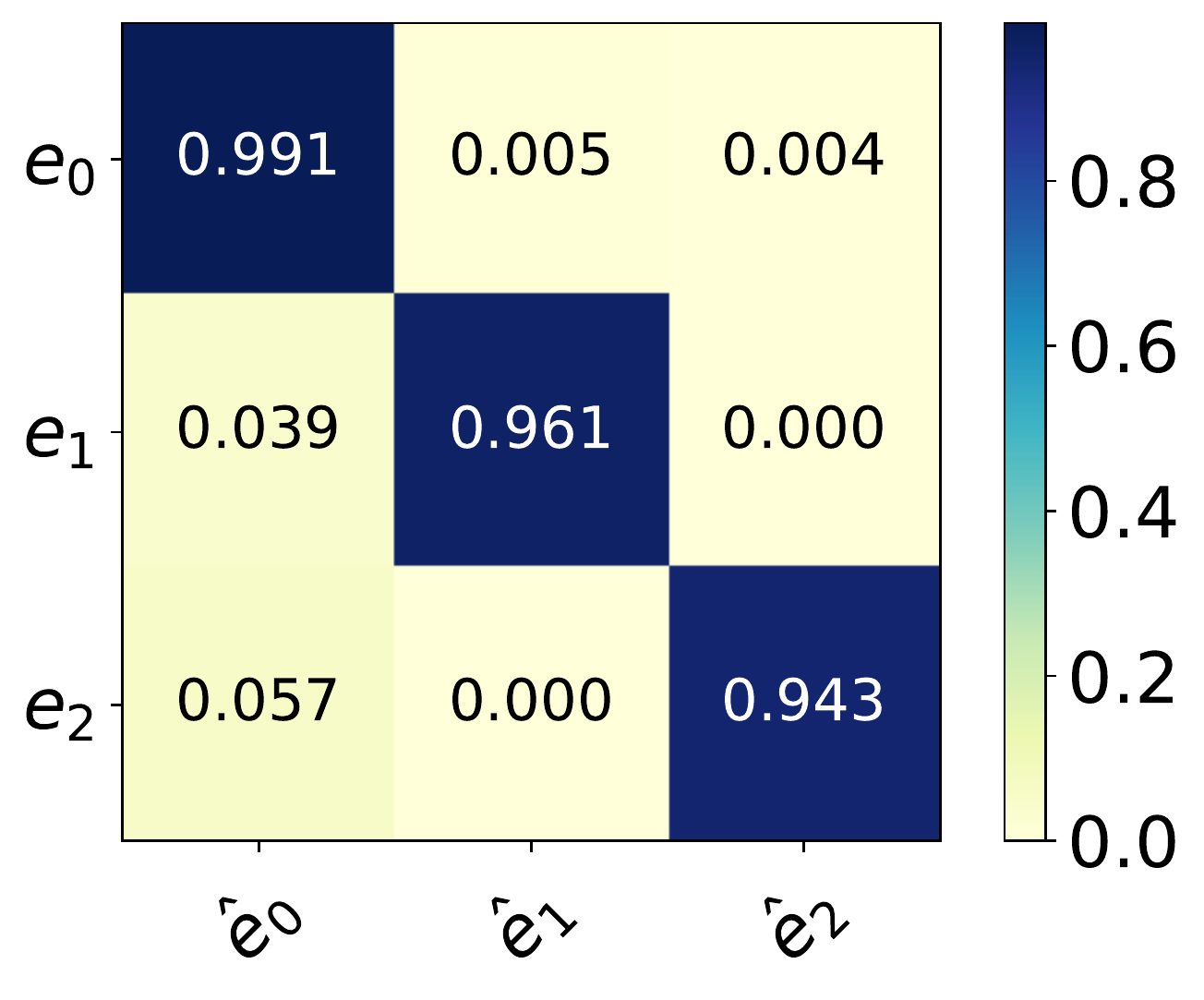}} 
            \hfil
        \caption{Confusion matrices of CRDNNs} \label{fig:fig6}
\end{figure*} 

As shown in Fig. \ref{fig:fig5}, the cost goes down to a certain level and then fluctuates if the regularization and drop-out method are used. Notice that the cost with anti-overfitting methods is higher since we use the regulation method: it does not mean the accuracy is worse than the one without anti-overfitting methods. Also, we add weight to the cost function. The weight can avoid a certain kind of error by recognizing. For instance, if the weight on loading is higher than the weight on the traveling process, the optimization process will take more attention to avoid the errors on loading rather than on the traveling process. Formally, see eq. \eqref{eq3}.

\begin{equation}
\begin{split}
    J(\theta) = \frac{1}{m}\sum \limits_{i=1}^{m}\sum \limits_{k=1}^{K}[-y_k^{(i)}\log((h_\theta x^{(i)})_k)\\\
    -(1-y_k^{(i)})\log(1-(h_\theta x^{(i)})_k) ]W_k\\\
    +\frac{\lambda}{2m}\left[\sum \limits_{j=1}^{32}\sum \limits_{k=1}^{32}(\theta_{j,k}^{(1)})^2+\sum \limits_{j=1}^{3}\sum \limits_{k=1}^{32}(\theta_{j,k}^{(2)})^2\right]
    \label{eq3}
\end{split}
\end{equation}

Obviously, compared to the cost function without regularization, the regularization might increase the total value of cost function. The $W_k$ denotes the weight of k state. In our case, we recommend setting the weight as
\begin{equation}
    \tilde{W} = {\left[ \begin{array}{ccc}
        1 & 4 & 7
    \end{array}\right]}^T
    \label{eq4}
\end{equation}

Since rectified linear unit (ReLu) has a constant gradient if the X$>$0, we use ReLu as activation function so that the calculation effort can be reduced and thereby converging or learning much faster.
The Hyper-parameters we used are shown in Table \ref{tab_2}.

\begin{table}[!ht]
	\centering
	\caption{Parameters of CRDNN}
	\begin{tabular}{l|cl}
	\hline \hline
	Hyper-parameters
		& Value \\ \hline
	Hyper-parameters
		& [9, 15, 25] \\
	Batch size
		& 128 \\
   	Initial learning rate (decay during learning)
		& 1x$10^{-4}$ \\
	Num filter conv1D
		& 10 \\
	Kernel size 
	    & 5 \\
	Num units 1$^{st}$ layer (RNN) 
	    & 32 \\
    Num units 2$^{nd}$ layer (RNN) 
        & 32 \\
    Num units 1$^{st}$ layer (DNN) 
        & 32 \\
    Num units 2$^{nd}$ layer (DNN)
        & 32 \\

	\hline \hline 
	\end{tabular}
	\label{tab_2}
\end{table}

Generally, we shall use the F1 micro average to evaluate and select the best suitable architecture.  Nonetheless, since we are going to implement an operation strategy based on the learning algorithm later, the F1 score alone does not indicate whether a result is easy to correct or not, so we also use confusion matrices to evaluate the results, see Fig. \ref{fig:fig6}, where the abscissa indicates the predicted value and the ordinate indicates the ground truth label. $e_0$, $e_1$, $e_2$ denote the travelling process, the loading process, and the unloading process appropriately. The F1 score is used as a subordinate criterion to select the better overall performance solution. Obviously, the CRDNN with two bidirectional LSTMs has the best performance, which is similar to our assumption, with an overall accuracy at 98.5\%. Compared to simple DNNs, CRDNN has an improvement of about 3\%. Bidirectional LSTMs make the decision using a relatively more prolonged-time period and can consider the data after the event so with no doubt it has better accuracy. The improvement compared to DNN is because LSTMs are good at dealing with long term problems so that we can use a larger window size to fit into CRDNNs. Another potential architecture is CRDNN with two LSTMs, which is only slightly worse than the one with bidirectional LSTMs but the training parameters are much fewer. An additional advantage of CRDNNs is that it has even fewer training parameters though it has a complicated architecture. Compared to the simple neural networks with two hidden layers and 128 units per layer, CRDNN with two layers of LSTMs has only 16,295 training parameters while the former has more than 30,000 training parameters, resulting in a much faster on-board calculation.

\section{EVALUATION OF THE METHODS}
\begin{figure}[!htbp]
    \centerline{\includegraphics[width=3.5in]{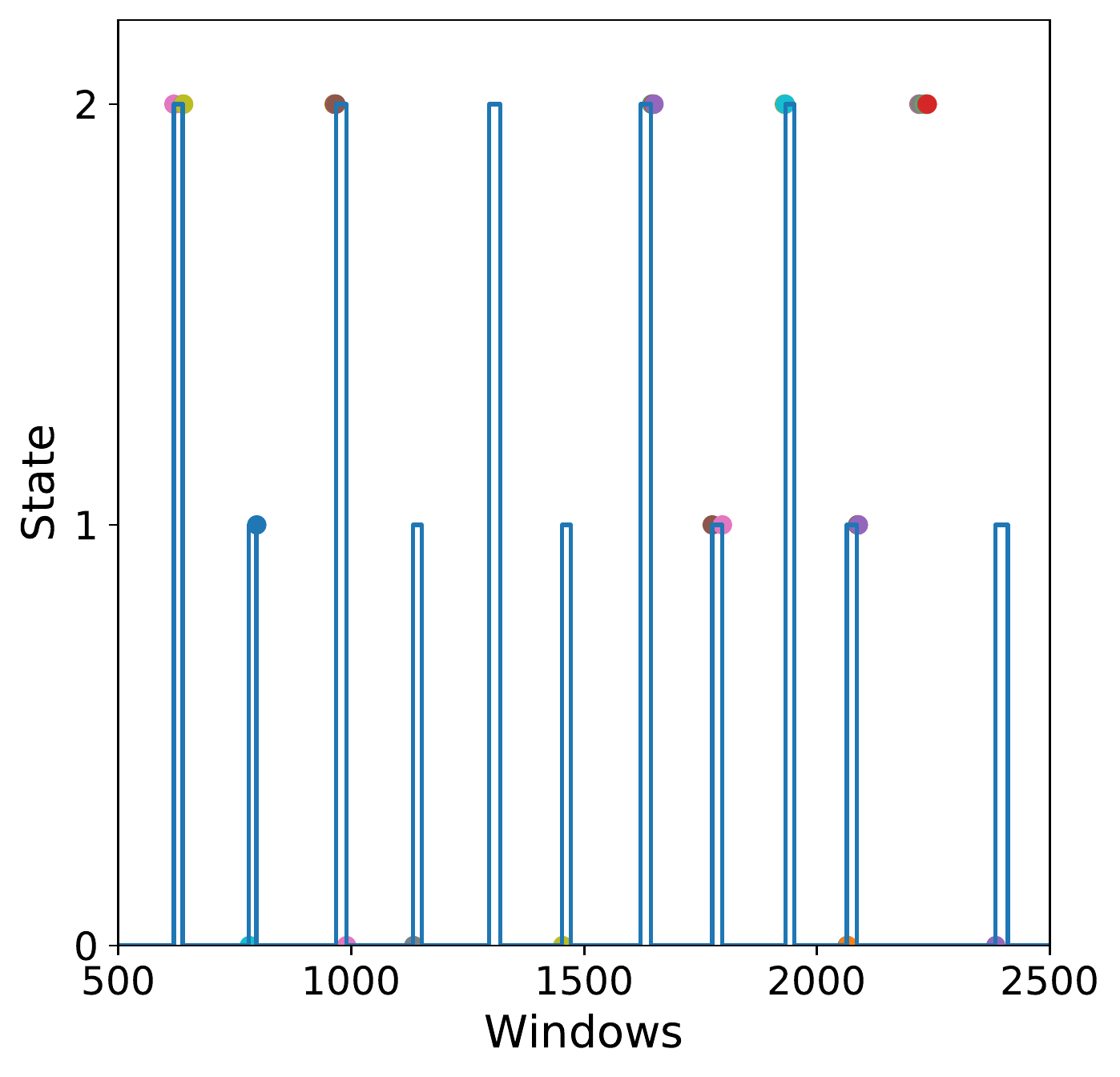}}
    \caption{Ground truth and prediction mistakes}
    \label{fig:fig7}
\end{figure}

As the results are shown in the last section, the improvement of test accuracy is almost stopped at 98.2\%. To further improve the prediction accuracy, we draw out the place where our algorithm has made a mistake.
As shown in Fig. \ref{fig:fig7}, we draw the ground truth and the mistakes made by CRDNN with two bidirectional LSTMs. The blue line denotes the ground truth label and the color points represent the place where CRDNN recognizes a different result as ground truth label and thus we say it makes a mistake.  Obviously, the mistakes mainly occur at the time when the machine changes its state from one to another. However, we can say that they are the states which are also controversial for humans to say whether the state should be loading, unloading, or travel since the features are vague in this region. When we further draw all the falsely recognized time windows, we found that almost all of the mistakes occur when the state is really fuzzy.

One exception is the windows around the 2250th window in Fig. 7, corresponding to the measurement data at 450 seconds, see Fig. 3. As mentioned before, the time windows at 450 seconds are the windows that we consciously mislabeled to the travel process. This proves that our CRDNN has a robust performance even if a few data is mislabeled. That is, the CRDNN with two bidirectional LSTMs accurately identified that the process is actually an unloading process rather than a travel process. Therefore, we believe that about 98\% is the best number since different engineers define the ground truth differently with their plausible reasons. Moreover, although a tiny number of mislabeling might harm the test accuracy. However, it cannot affect the prediction performance of the CRDNN.

\section{CONCLUSION AND OUTLOOK}
In this paper, we have shown that CRDNN with bidirectional LSTMs has the best performance to detect the truck loading cycles, and the CRDNN with 2 LSTMs has the best performance-cost ratio if primary torque control concept is used. Because we use an offline learning strategy and the forward propagation is much faster than backward propagation, this method will not take up too much computational effort. By considering a period of 5 seconds, the test accuracy reaches 98.2\%, and it never mistakes the loading process with the unloading process or vice versa, which makes the operation strategies easily to be implemented. Also, since we have a large dataset, a tiny mislabel could not harm the real performance of the CRDNNs. It is also worthy to point out, although CRDNN has only increased the test accuracy by 3\%, it increases the most challenging 3\% by successfully detecting the state of the data gathered when the drivers did not operate well. As a result of successful detection of truck loading process, our primary torque controlled wheel loader can increase its efficiency up to 9\% due to regeneration process. 
The dataset and the code in this paper are also published on our Github website.


\subsection{Outlook}
Although the proposed deep learning algorithm can successfully handle the time series problem so as we can further improve the operation strategy of mobile machines, this method is challenging to predict the next process of mobile machines since truck loading cycles are quite different from the others. However, the Pandora’s box has not been fully opened yet. To achieve autonomous working mobile machines or further increasing work efficiency using artificial intelligence, the advantages of image recognition must be taken into account. With a camera, we can not only detect the current working process of mobile machines but also predict its intention. Consequently, power management can make better preparation.

In our next paper, we are going to introduce the two-dimensional image detection methods to detect mobile machines. The method proposed in this paper also provides the relevant information so that we can fuse them to produce even higher accurate classification or prediction results. A regret is that we did not spend much time optimizing the CRDNNs to detect truck loading processes due to the limited time. With further optimization, at least the training parameters can be reduced so that an even faster CRDNN can be expected.


\bibliography{Literature_IEEE.bib}{}
\bibliographystyle{IEEEtran}

\end{document}